\documentclass{article}
 \usepackage[hypertex]{hyperref}  

\usepackage{graphicx}
\usepackage{amsmath}
\usepackage{amsfonts}
\usepackage{amssymb}
\usepackage{slashed}
\providecommand{\N}[1]{\ensuremath{\tilde{\chi}^o_#1}}

\providecommand{\GeV}{\,\,{\rm{GeV}}}

\providecommand{\fb}{\,\,{\rm{fb}}}

\providecommand{\herwig}{{\tt HERWIG}}

\providecommand{\max}{{\tt{max}}\ }

\providecommand{\etal}{{\emph{et.al}}}

\usepackage{color}

\begin{document}

\author{Alan J. Barr$^{1}$\thanks{a.barr@physics.ox.ac.uk}\ and Alex Pinder$^{1}$\thanks{a.pinder1@physics.ox.ac.uk} \ and Mario Serna$^{2}$\thanks{mariojr@alum.mit.edu} \\
\small $^{1}$ Department of Physics, Denys Wilkinson Building,
University of Oxford, Keble Road, Oxford OX1 3RH, United Kingdom\\
\small $^{2}$ Rudolf Peierls Centre for Theoretical Physics,
University of Oxford, Keble Road, Oxford, OX1 3NP, United Kingdom \\ } \normalsize

\title{Precision Determination of Invisible-Particle Masses at the CERN LHC: II}
\maketitle

\begin{abstract}
We further develop the constrained mass variable techniques to determine the mass scale of invisible particles pair-produced at hadron colliders.  We introduce the constrained mass variable $M_{3C}$ which provides an event-by-event lower bound and upper bound to the mass scale given the two mass differences between the lightest three new particle states.  This variable is most appropriate for short symmetric cascade decays involving two-body decays and on-shell intermediate states which end in standard-model particles and two dark-matter particles.  An important feature of the constrained mass variables is that they do not rely simply on the position of the end point but use the additional information contained in events which lie far from the end point.  To demonstrate our method we study the supersymmetric model SPS 1a.  We select cuts to study events with two $\N{2}$ each of which decays to $\N{1}$, and two opposite-sign same-flavor (OSSF) charged leptons through an intermediate on-shell slepton.  We find that with $300 \fb^{-1}$ of integrated luminosity the invisible-particle mass can be measured to $M_{\N{1}}=96.4 \pm 2.4$ GeV.  Combining fits to the shape of the $M_{3C}$ constrained mass variable distribution with the $\max m_{ll}$ edge fixes the mass differences to $\pm 0.2$ GeV.
\end{abstract}

\section{Introduction}

If dark matter is produced at a hadron collider, the likely signature will be missing transverse momentum.
In previous papers, we have introduced
the constrained mass variable $M_{2C}$ \cite{Ross:2007rm} \cite{Barr:2008ba} as a means to determine the mass of the dark matter.
 The main concept behind the constrained mass variable $M_{2C}$  is that after studying several kinematic quantities we may have well determined the mass difference between two particle states but not the mass itself. We then incorporate these additional constraints in the analysis of the events. We check each event to test the lower bounds and upper bounds on the mass scale that still satisfies the mass difference and the on-shell conditions for the assumed topology.  Because the domain over which we are minimizing contains the true value for the mass, the end points of the lower-bounds' and upper-bounds' distributions  give the true mass.

\begin{figure}
\centerline{\includegraphics[width=4in]{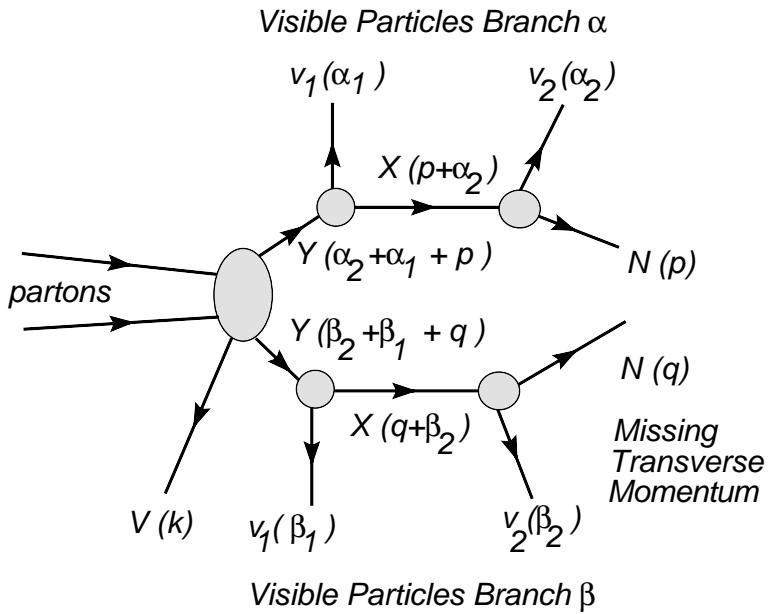}}
\caption{\label{FigEventTopologyTovey} Events where the new state $Y$ is pair produced and in which each $Y$ decays
through a two-body decay to a massive new state $X$ and a visible state $1$ and then where $X$ subsequently decays to a massive state $N$ invisible to the detector
and visible state $2$.  All previous decay products are grouped into the upstream transverse momentum, $k$. }%
\end{figure}
In previous studies \cite{Ross:2007rm}\cite{Barr:2008ba} the constrained mass variable $M_{2C}$ was introduced and studied.  The constrained mass variable $M_{2C}$  assumes a new state $Y$ decays to visible particles $(1)$ and $(2)$ and $N$ through a three-body decay in which the mass difference is given by the end point of the invariant-mass distribution of the two visible states on a branch
 \begin{equation}
  \max m^2_{12} = (M_Y - M_N)^2.
   \label{EqThreeBodyDecayEdge}
 \end{equation}
In the situations we study, visible states (1) and (2) are leptons so we will refer to this edge as $\max m_{ll}$.
We also note that the $M_{2C}$ may also compliment or cross-check other mass-determination techniques \cite{Bachacou:1999zb,phdthesis-lester,Allanach:2000kt,Gjelsten:2004ki,Bisset:2008hm,Lester:2006yw,Gjelsten:2006tg,Cheng:2007xv,Nojiri:2007pq,Cheng:2008mg,Cho:2007qv,Cho:2007dh,Barr:2007hy}
which may somehow otherwise determine the mass difference but may have a remaining ambiguity on the mass scale.
The purpose of this paper is extending the constrained mass variable to the case
with three new on-shell states as depicted in Fig.~\ref{FigEventTopologyTovey}.
With an on-shell intermediate state, the kinematic edge from the end point of the invariant-mass distribution of the visible states (1) and (2) on a branch gives the relationship
 \begin{equation}
 \max m^2_{12} = \frac{(M_Y^2 -M_X^2)(M_X^2 -M_N^2)}{M_X^2}.
 \label{EqTwoBodyDecayEdge}
 \end{equation}
Each event now satisfies an additional set of on-shell constraints so the events should contain more information.
Because Eq.~(\ref{EqThreeBodyDecayEdge}) does not give the mass difference and because the $M_{2C}$ variable does not use the additional information available from having three on-shell states in each event,  then a better variable with which to find the mass scale likely exists by incorporating this missing information in the extremization.

In this paper we introduce a constrained mass variable more appropriate for this case, one with an on-shell intermediate state, which we will call $M_{3C}$.
The variable $M_{3C}$ differs from $M_{2C}$ in that we assume an on-shell intermediate state $X$ connects the two visible decay products so there are three new states and two relevant mass differences.
We structure the paper around a case study of the supersymmetry benchmark point SPS 1a \cite{Allanach:2002nj}.
In this study, the three new states are identified as  $Y=\N{2}$, $X=\tilde{l}$ and $N=\N{1}$.  The visible particles leaving each branch are all opposite-sign same-flavor (OSSF) leptons ($\mu$ or $e$).  This allows us to group hadronic activity into the vector $k$ identified as upstream transverse momentum (UTM).

The paper is structured as follows:
Section~\ref{SecM3CIntro} introduces the definition of $M_{3C}$. At this stage we assume we know the two mass differences, an assumption which will be justified later in the paper.  Section~\ref{SecM3CDependence} discuses the dependence of $M_{3C}$ on complications from combinatorics, large UTM, missing transverse-momentum ($\slashed{P}_T$) cuts, parton distributions, and energy resolution.   Section~\ref{SecM3CPerformance} applies $M_{3C}$ variables to \herwig\ data from the benchmark supersymmetry spectrum SPS 1a.  Section~\ref{SecMassDifferences} shows how combining the edge from Eq.~(\ref{EqTwoBodyDecayEdge}) with $M_{3C}$ one also finds the two mass differences $M_Y-M_N$ and $M_{X}-M_N$.
Finally in Sec.~\ref{SecM3CConclusions} we summarize the papers's contributions.

\section{Introducing $M_{3C}$}
\label{SecM3CIntro}
We will now introduce the definition of $M_{3C}$ and its relationship to previous mass-shell techniques (MST).

\subsection{Definition of $M_{3C}$}

The upper-bound and lower-bound
on the mass of the third lightest new particle state in the symmetric decay chain are the constrained mass variables $M_{3C,LB}$ and $M_{3C,UB}$.
This variable applies to the symmetric, on-shell intermediate state, topology from Fig.~\ref{FigEventTopologyTovey} which depicts
two partons that collide and produce some observed UTM with four momenta $k$ and an on-shell, pair-produced new state $Y$.
On each branch, $Y$ decays to on-shell intermediate particle state $X$ and a visible particle $v_1$ with masses $M_{X}$ and $m_{v_1}$. Then $X$ decays to the dark-matter particle $N$ and visible particle $v_2$ with masses $M_N$ and $m_{v_2}$.
The  four-momenta of $v_1$, $v_2$ and $N$ are respectively $\alpha_1$, $\alpha_2$ and $p$ on one branch and $\beta_1$, $\beta_2$ and $q$ in the other branch.
The missing transverse momenta $\slashed{P}_T$ is given by the transverse part of $p+q$.

We initially assume that we have measured the mass differences from other techniques.  For an on-shell intermediate state, there is no single end point that gives the mass difference. The short decay chain gives a kinematic end point $\max m_{12}$ described in Eq.~(\ref{EqTwoBodyDecayEdge}) that constrains a combination of the squared mass differences.  Unless two of the states are nearly degenerate, the line with constant mass differences lies very close to the surface given by Eq.~(\ref{EqTwoBodyDecayEdge}).  The two mass differences are often tightly constrained in other methods.
The mass differences are constrained to within $0.3$ GeV from studying long cascade decay chains where one combines constraints from several end points of different invariant-mass combinations \cite{Gjelsten:2004ki}.
In principle mass differences may also be found by using $M_{T2}$ with different assignments of upstream transverse momentum and missing transverse momentum as described in Ref.~\cite{Serna:2008zk}.
After initially assuming that we know the mass difference, we show that our technique can also find the mass differences.
The $M_{3C}$ distribution shape is a function of both the mass scale and mass differences.
We can constrain both the mass differences and the mass scale by fitting the $\max m_{12}$ edge constrains
and the ideal $M_{3C}(M_N,\Delta M_{YN},\Delta M_{XN})$ distribution shapes to the observed $M_{3C}(\Delta M_{YN},\Delta M_{XN})$.  To find all three parameters from this fit, we will take $M_{N}$, $\Delta M_{YN}$, and $\Delta M_{XN}$  as independent variables.

For this first phase of the analysis, let us assume the mass differences are known.
For each event, the variable $M_{3C,LB}$ is the minimum value of the mass of $Y$ (third lightest state) after minimizing over the unknown division of the missing transverse energy $\slashed{P}_T$ between the two dark-matter particles $N$:
 \begin{eqnarray}
   m^2_{3C,LB}(\Delta M_{YN},\Delta M_{XN}) &= & \min_{p,q}\  (p+\alpha_1+\alpha_2)^2 \label{EqM3CTop} \\
{\rm{Constrained}\ \rm{to}}  & &  \nonumber \\
    (p+q)_T & = & \slashed{P}_T \label{EqM3Cc1} \\
    \sqrt{(\alpha_1+\alpha_2+p)^2} -     \sqrt{(p^2)}& = &\Delta M_{YN} \\
        \sqrt{(\alpha_2+p)^2} -     \sqrt{(p^2)} & = & \Delta M_{XN} \\
    (\alpha_1+\alpha_2+p)^2 & =&  (\beta_1+\beta_2+q)^2 \\
        (\alpha_2+p)^2 & = & (\beta_2+q)^2 \\
                p^2 & = & q^2 \label{EqM3Cc7}
 \end{eqnarray}
where $\Delta M_{YN} = M_Y - M_N$ and $\Delta M_{XN} = M_X - M_N$.
There are eight unknowns in the four momenta of $p$ and $q$ and seven equations of constraint. Likewise we define $M_{3C,UB}$ as the maximum value of $M_Y$ compatible with the same constraints.   We discuss how to numerically implement this minimization and maximization in appendix~\ref{SecNumericallyCalculatingM3C}.  Because the true $p$ and $q$ are within the domain over which we are minimizing (or maximizing), the minimum (maximum) is guaranteed to be less than (greater than) or equal to $M_Y$.
Just as with $M_{2C}$, events with $M_{3C}$ (upper or lower) near the end-point are nearly reconstructed \footnote{In-principle we expect to find some events close to the end point.  In practice it may be more challenging with $M_{3C}$ than with $M_{2C}$.  The $M_{3C}$ variable is more sensitive to energy-resolution errors so the sharp cut off at the correct mass becomes a sloping gradual tail as seen in Fig~\ref{FigM3CEnergyResolution} which obscures events are near the end-point.
The $M_{2C}$ end point is better for the purpose of reconstruction.  Although fewer events will be near this end-point, the likelihood of energy resolution errors to completely mess up the results is less. This reconstruction near the endpoint is being exploited to study spin with the $M_{T2}$ assisted on-shell (MAOS) reconstruction \cite{Cho:2008tj}.}.  We provide a proof of the uniqueness of such a reconstruction in appendix \ref{AppendixUniquenessOfReconstruction}.

\begin{figure}
\centerline{\includegraphics[width=5in]{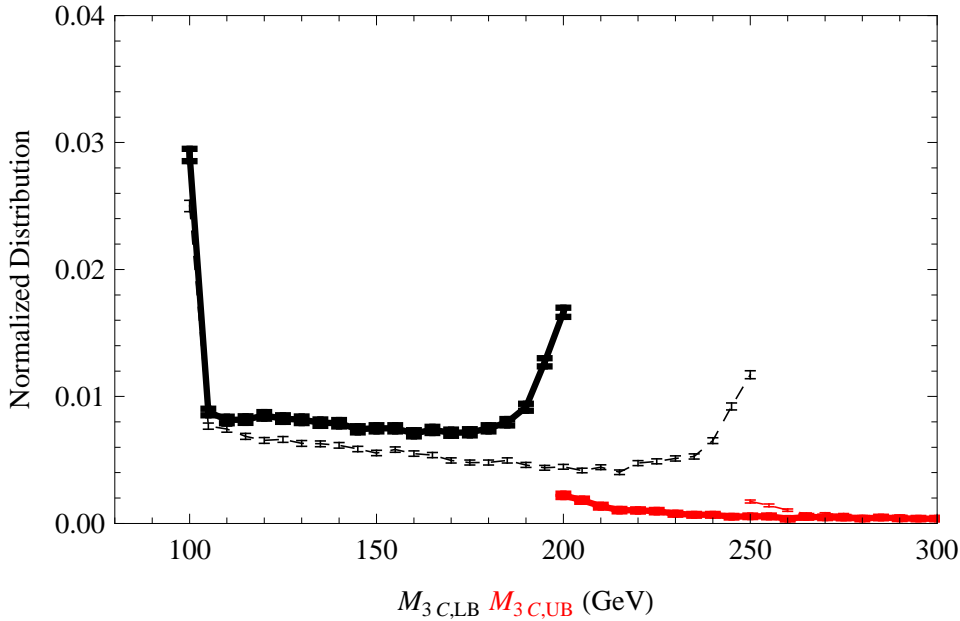}}
\caption{\label{FigM3CIdeal150100} Ideal $M_{3C,LB}$ and  $M_{3C,UB}$ distribution for 25000 events in two cases both sharing $\Delta M_{YN}=100$ GeV and $\Delta M_{XN}=50$ GeV. The solid, thick line shows $M_Y=200$ GeV, and the thin, dashed line shows $M_Y=250$ GeV.}
\end{figure}

Figure \ref{FigM3CIdeal150100} shows an ideal $M_{3C,LB}$ and $M_{3C,UB}$ distributions for $25000$ events in two cases both sharing $\Delta M_{YN}=100$ GeV and $\Delta M_{XN}=50$ GeV. The dashed lines represent the distributions from events with $M_Y=250$ GeV, and the solid lines represent the distributions from events with $M_{Y}=200$ GeV.  One can clearly see sharp end points in both the upper-bound and lower-bound distributions that give the value $M_Y$. The upper-bound distribution is shown in lighter line (red online).

We expect an event to better constrain the mass scale
if we are given additional information about that event.
In comparison to $M_{2C}$ where $Y$ decays directly to $N$,
here there is an on-shell intermediate state $X$.
The extra state $X$ and information about its mass difference $\Delta M_{XN}$ enables $M_{3C}$ to make an event-by-event bound on $M_Y$ stronger than in the case of $M_{2C}$.  We will see that this stronger bound is partially offset by greater sensitivity to errors in momentum measurements.

The variable $M_{3C}$, like other variables we have discussed $M_{2C}$, $M_{T2}$ $M_T$ and $M_{CT}$, is invariant under longitudinal boosts of its input parameters.
We can understand this because all the constraint equations are invariant under longitudinal boosts.  The unknown $p$ and $q$ are minimized over all possible values fitting the constraints so changing frame of reference will not change the extrema of the Lorentz invariant quantity $(p+\alpha)^2$.

\subsection{Relation to other Mass-Shell Techniques}

There are several tools being studied to determine the mass of the dark-matter particle in a model-independent manner.  We now discuss $M_{3C}$ in the context of recent papers.

The variable $M_{3C}$ is a constrained mass variable where we find minimum and maximum mass allowed by an event given the combination of on-shell requirements for an assumed topology, the observed missing transverse momentum, and mass differences which may extracted from kinematic end-points or other techniques.
The variable $M_{2C}$ \cite{Ross:2007rm}\cite{Barr:2008ba} was the first example of a constrained mass variable.  This paper's $M_{3C}$ is the next example.

The kinematic variable $M_{T2}$ \cite{Lester:1999tx,Barr:2003rg} (in addition to the original definition) can be seen as the lower-mass boundary of the region with these minimal kinematic constraints and an assumed mass for the dark-matter particle \cite{Cheng:2008hk}.
$M_{T2}$ can be applied to subsystems of an assumed topology as outlined in \cite{Serna:2008zk} and further studied more recently in \cite{Burns:2008va}.
The variable $M_{3C}$ does not use $M_{T2}$ as a calculation tool,
but the variable $M_{3C}$ can also been seen as the boundary of
the `non-minimal'\footnote{The term `non-minimal' is because we assume the more complex topology involving an on-shell intermediate state.} kinematic constraint mass region \cite{Cheng:2008hk} for the topology with three on-shell new states shown in Fig~\ref{FigEventTopologyTovey}.

Mass-shell techniques (MSTs) require the consistency of each event with missing transverse momentum and with on-shell requirements for an assumed topology. 
Mass-shell techniques\cite{Bisset:2008hm} encompasses the polynomial methods described in Ref.~\cite{Burns:2008va} and encompasses the minimal kinematic constraints method described in Ref~\cite{Cheng:2008hk} as well as many others reviewed in our earlier work\cite{Ross:2007rm}\cite{Barr:2008ba}.

Since the publication of Ref.~\cite{Ross:2007rm}, the combination of
 mass-shell constraints with kinematic end-point constraints has stated being referred to  hybrid mass-shell techniques\cite{Nojiri:2007pq,Burns:2008va}. 
The MST most closely associated with $M_{3C}$ is that of
Cheng, Gunion, Han, Marandella, McElrath (CGHMM) \cite{Cheng:2007xv} which describe counting solutions at assumed values for the mass for $Y$, $X$, and $N$.
By incorporating a minimization or maximization, we enhance CGHMM's approach because we have a variable whose value changes slightly with slight changes of the inputs instead of the binary on-off that CGHMM have with the existence of a solution\footnote{We am grateful to Chris Lester for pointing out to me the importance of this feature.}.
As a hybrid technique, we also incorporate knowledge of the added information from other measurements which accurately determine the mass differences.
Finally, the quantity $M_{3C}$ can form a distribution whose shape tells us information about the masses.
Because for most events there is only one ``turn-on" point below $M_Y$, the distribution $M_{3C,LB}$ is very similar to the derivative of the Fig.~8 of CGHMM\cite{Cheng:2007xv} to the left of their peak and $M_{3C,UB}$ is similar to the negative of the derivative to the right of their peak.
They differ in that there may be multiple windows of solutions; also  CGHMM's Fig.~8 is not exactly along the line of fixed mass differences; and the effect of backgrounds and energy resolution are dealt with differently.  

We also hope to show that the use of the distribution's shape enables us
to exploit the essentially non-existent dependence of the distributions on the unknown collision energies and incorporate the dependency on UTM directly.  This diminishes the dependence of the measurement on the unknown model while still allowing us to exploit the majority of the distribution shape in the mass determination.

After studying previous MSTs, we were tempted to use Bayes' theorem with  a parton distribution function as a likelihood function as was done in Goldstein and Dalitz \cite{Goldstein:1993mj} and Kondo, Chikamatsu, Kim \cite{Kondo:1993in} (GDKCK).
They used the parton distribution function to weight the different mass estimates of the top-quark mass ($M_Y$ in our topology).
We found that such a weighting leads to a prediction for $M_Y$ much smaller than the true value.  This can be understood because the parton distributions make collisions with smaller center-of-mass energies (small $x$) more likely, therefore the posterior will prefer smaller values of $M_Y$ which are only possible for smaller values of $x$.
Only if one includes the cross section for production, i.e. the likelihood of the event existing at all, in the Bayes likelihood function will we have the appropriate factor that suppresses small values of $x$ and therefore small values of $M_Y$.  This balance therefore leads to the maximum likelihood (in the limit of infinite data) occurring at the correct $M_Y$.  Unfortunately, inclusion of the magnitude of the cross section introduces a large model dependence.
In the case of the top-quarks mass determination, the GDKCK technique gives reasonable results. This is because they were not scanning the mass scale, but rather scanning $\chi_{Y}$ (the top-quark mass) while assuming $\chi_N=M_N=0$ and $\chi_X=M_X=M_W$.
The likelihood of solutions as one scans $\chi_{Y}$ rapidly goes to zero below the true top-quark mass $M_{top}$.
The parton distribution suppresses the likelihood above the true $M_{top}$.
The net result gives the maximum likelihood near the true top-quark mass but suffers from a
systematic bias \cite{Raja:1996vz}\cite{Raja:1997qs} that must be removed by modeling \cite{Brandt:2006uc}.

\section{Factors for Successful Shape Fitting}
\label{SecM3CDependence}

One major advantage of using the $M_{3C}$ distribution (just as the $M_{2C}$ distribution \cite{Ross:2007rm}\cite{Barr:2008ba}) is that the bulk of the relevant events are used to determine the mass
and not just those near the end point.  To make the approach mostly model independent, we study on what factors the distributions shape depends.
We show that there is a strong dependence on UTM,
but because the UTM distribution can be measured from the data this
does not increase the model dependency.
We show there is no numerically significant dependence on the collision energy which is distributed according to the parton distribution functions.
This makes the distribution shape independent of the production cross section and the details of what happens upstream from the part of the decay chain that we are studying.
We model these effects with a simple Mathematica Monte Carlo event generator
assuming $M_Y=200$ GeV, $M_X=150$ GeV, and $M_N=100$ GeV.

\begin{itemize}
\item {\bf{Effect of Combinatorics Ambiguities}}
\end{itemize}
Just as in the topology in  Refs~\cite{Ross:2007rm}\cite{Barr:2008ba} studied earlier, where $\N{2}$ decays via a three body decay, the branch assignments can be determined by either distinct OSSF pairs or by studying which OSSF pairs have both $m_{12} \leq \max m_{12}$.
In $90 \%$ of the events, there is only one combination that satisfy $m_{12} \leq \max m_{12}$.  This allows us to know the branch assignment of $95 \%$ of the four-lepton events without ambiguity.

\begin{figure}
\centerline{\includegraphics[width=3.1in]{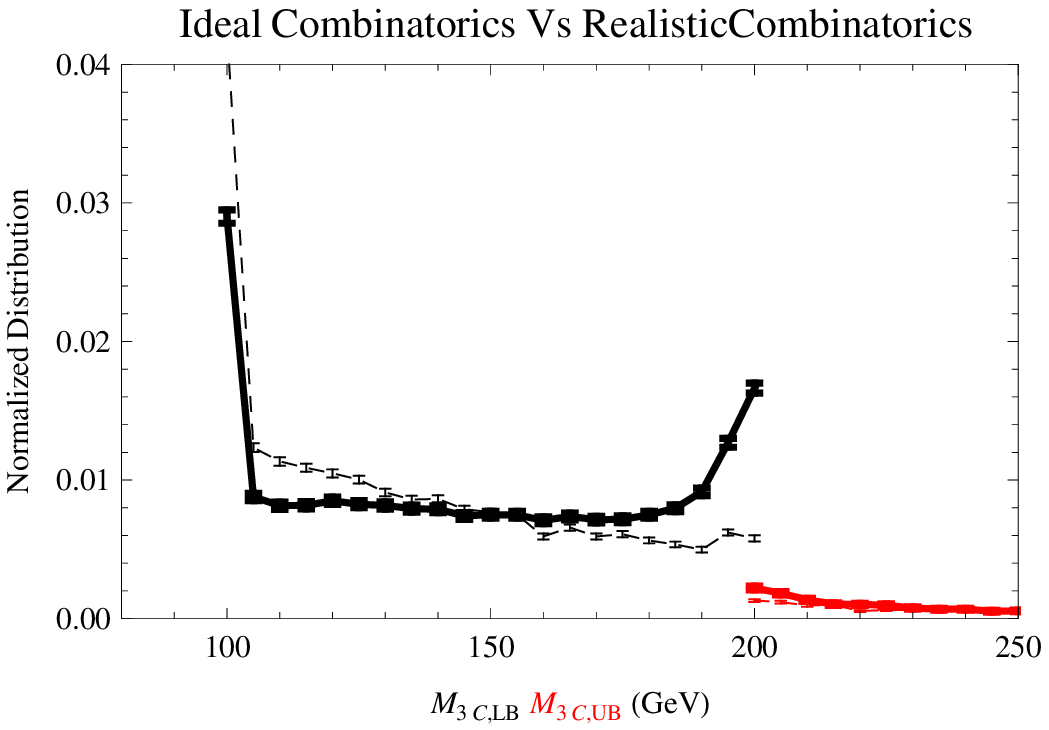}
\includegraphics[width=3.1in]{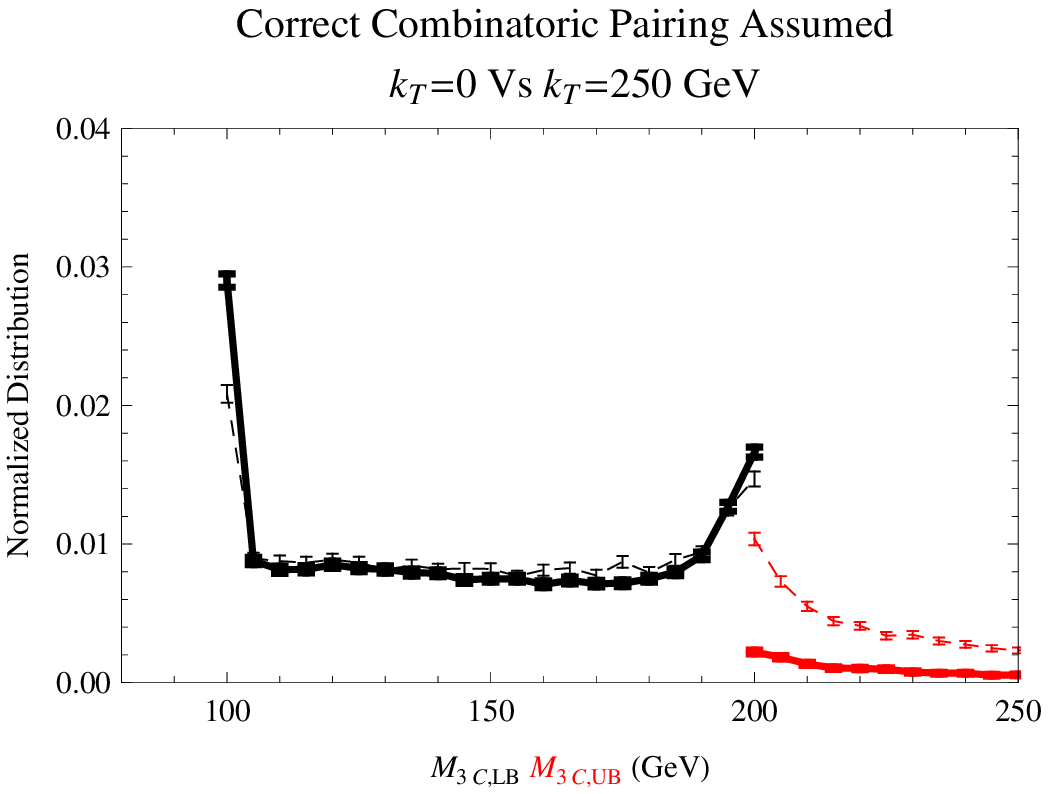}}
\caption{\label{FigM3CCombinatoricsDemo}\label{FigM3CUTMDistribution} (Left) The $M_{3C}$ distributions before (solid) and after (dashed) introducing the combinatoric ambiguity.  (Right) The $M_{3C}$ distributions with  and without UTM.
The no UTM case ($k_T=0$) is shown by the solid line; the large UTM case with $k_{T}=250$ GeV is shown by the dashed line.
}
\end{figure}

Unlike the three-body decay case, the order of the two leptons on each branch matters. The intermediate mass $M_X^2 = (\alpha_2 + p)^2$ depends on $\alpha_2$ and does not depend on $\alpha_1$.
To resolve this ambiguity we consider the four combinations that preserve the branch assignment but differ in their ordering.  The $M_{3C,LB}$ for the event is the minimum of these combinations.  Likewise the $M_{3C,UB}$ is the maximum of these combinations. As one expects, Fig.~\ref{FigM3CCombinatoricsDemo} (Left) shows how the combinatorics ambiguity degrades the sharpness of the cut-off at the true mass.  Not all applications share this ambiguity; for example in top-quark mass determination (pair produced with $Y=$ top quark, $X=W^\pm$, $N=\nu$) the $b$-quark-jet marks $\alpha_1$ and the lepton marks $\alpha_2$.

\begin{itemize}
\item {\bf{Effect Large Upstream Transverse Momentum}}
\end{itemize}

In a similar behavior to $M_{2C}$, the distributions of the variable $M_{3C}$
show a strong dependence on large UTM.
In our case study this is identified as the combination of all the hadronic activity.
Figure~\ref{FigM3CUTMDistribution} (Right) shows the stronger upper-bound cut-off in the presence of large UTM.  Unlike $M_{2C}$, in $M_{3C}$ with $k_T=0$ we still have events with nontrivial upper-bound values.

We also tested the distribution for different values of $k^2$.  In Fig.~\ref{FigM3CUTMDistribution} (Right) we fixed $k^2=(100 \GeV)^2$.  We also performed simulations with $k^2=(500 \GeV)^2$ and found the difference of the two $M_{3C}$ distributions consistent with zero after 15000 events.  In other words, the distribution depends mostly on $k_x$ and $k_y$ and appears independent of $k_0$.

\begin{itemize}
\item {\bf{The Effects of Detector Energy Resolution}}
\end{itemize}
\label{sec:detector}

\begin{figure}
\centerline{
\includegraphics[width=3.1in]{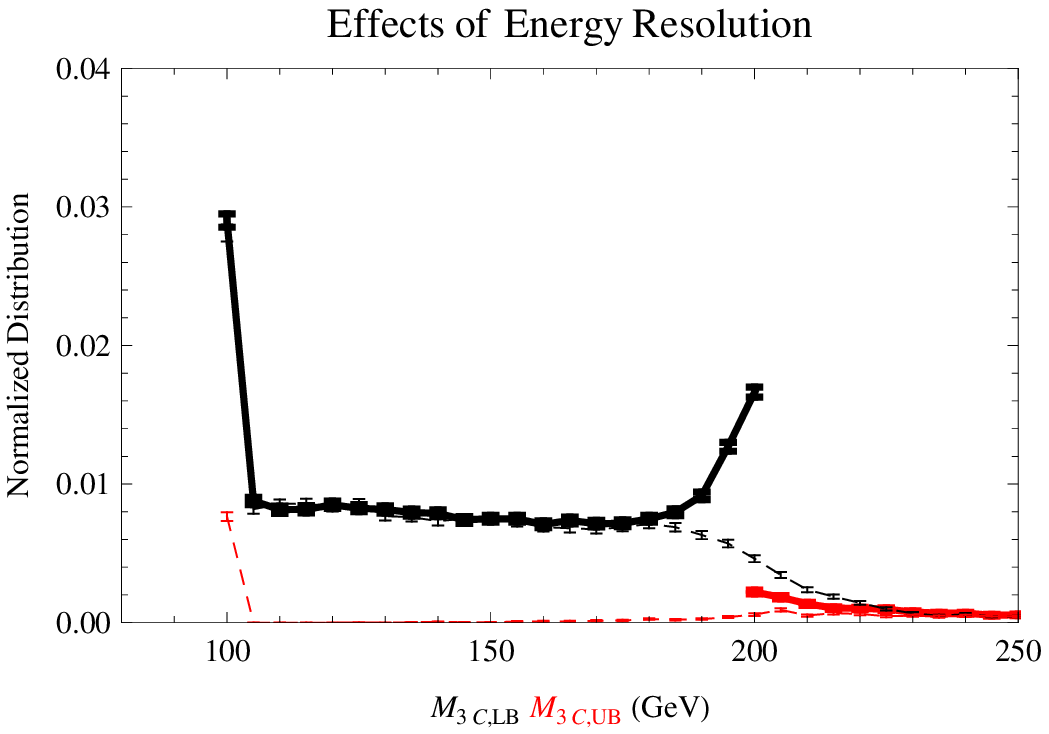}
\includegraphics[width=3.1in]{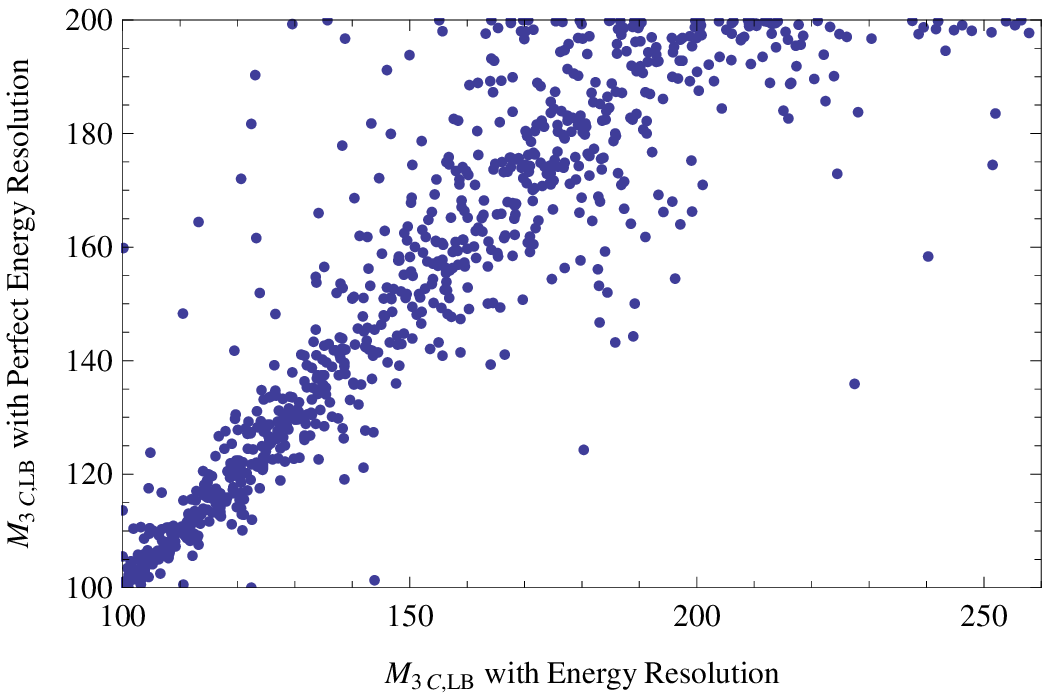}}
\caption{\label{FigM3CEnergyResolution}The effect of energy resolution on the $M_{3C}$ distribution.  (Left) The dotted line shows the energy resolution has washed out the sharp cut-off.
(Right) $M_{3C,LB}$ with perfect energy resolution plotted against the result with realistic energy resolutions.
}
\end{figure}

Compared to $M_{2C}$, the information about the extra states gives a stronger set of bounds.  Unfortunately, the solution is also more sensitive to momentum measurement error. We model the finite energy resolution by scaling the four-vector with gaussian centered around $1$ with the following widths which are similar to the expected resolutions of the ATLAS\cite{AtlasTDR} and CMS\cite{CMSTDR} detectors.
\begin{eqnarray}
 \frac{\delta E_e}{E_e} & = & \frac{0.1}{\sqrt{E_e}} + \frac{0.003}{E_e} + 0.007 \label{EqDetectorEnergyResolutionElectron} \\
 \frac{\delta E_\mu}{E_\mu} & = & 0.03  \label{EqDetectorEnergyResolutionMuon}\\
 \frac{\delta E_H}{E_H} & = & \frac{0.58}{\sqrt{E_H}} + \frac{0.018}{E_H} + 0.025.
 \label{EqDetectorEnergyResolutionHadron}
\end{eqnarray}
The hadronic energy-resolution error, which is larger  than the leptonic energy-resolution error, will increase the uncertainty in the missing transverse momentum.

Figure~\ref{FigM3CEnergyResolution} shows the effect of realistic leptonic energy
resolution for the case $k_T=0$ on the $M_{3C}$ distribution.
On the left we show the energy resolution (dashed line) compared to
 the perfect energy resolution (solid line).  The energy resolution washes out the sharp cut-off.
On the right we show $M_{3C,LB}$ with perfect energy resolution plotted against the result with realistic energy resolution.
This shows that the cut-off is strongly washed out because the events with $M_{3C}$ closer to the true value of $M_Y$ ($200$ GeV in this case) are more sensitive to energy resolution than the events with $M_{3C}$ closer to $\Delta M_{YN}$.  The peak in the upper-bound distribution at $M_{3C,UB}=100$ GeV comes from events that no longer have solutions after smearing the four-momenta.

Because the energy resolution affects the distribution shape, its correct modeling is important.  In the actual LHC events the
 $\slashed{P}_T$ energy resolution will depend on the hadronic activity in the events being considered.
Two events with the same $k_T=0$ may have drastically different $\slashed{P}_T$ resolutions.
Modeling the actual detector's energy resolution for the events used is important to predict the set of ideal distribution shapes which are compared against the low-statistics observed data.


We do not consider lepton-isolation cuts.  As described in Ref~\cite{Barr:2008ba}, including such effects changes the distribution shape and decreases the statistics available with which to form the distributions.
For example a lepton isolation cut of $\Delta R = \sqrt{\Delta \phi^2 + \Delta \eta^2} > 0.1$ excludes about $3\%$ of the distribution events or using $\Delta R > 0.3$ excludes about $20\%$ of the distribution events.  As lepton-isolation uncertainties and energy-resolution uncertainties are detector-specific, it would be interesting to study the distribution shape specific to different detectors.

\begin{itemize}
\item {\bf{Parton Distribution Function Dependence}}
\end{itemize}

For a mostly model-independent mass-determination technique, we would like to have a distribution that is independent of the specific production mechanism of the assumed event topology.
The parton distributions determine the center-of-mass energy $\sqrt{s}$ of the
hard collisions, but the cross section depends on model-dependent couplings and parameters.  The events we consider may come from production of different initial states (gluons or squarks) but end in the assumed decay topology.
The $M_{3C}$ distribution, like the $M_{2C}$ distribution, shows very little dependence on the underlying collision parameters or circumstances.

\begin{figure}
\centerline{
\includegraphics[width=3.1in]{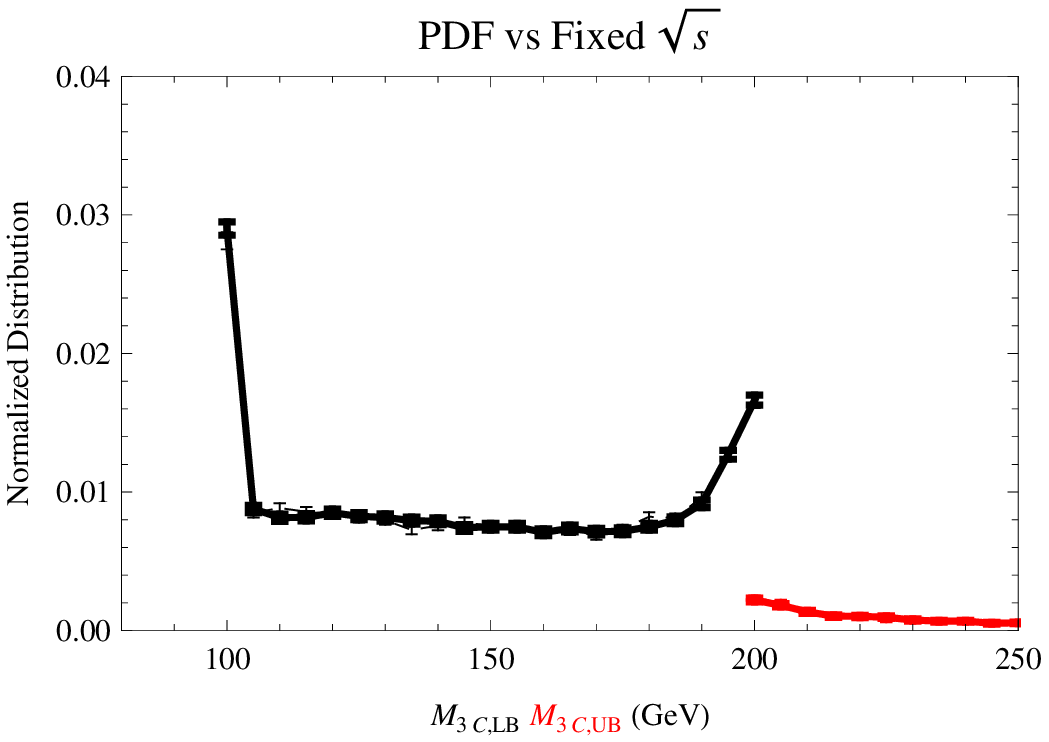}
\includegraphics[width=3.1in]{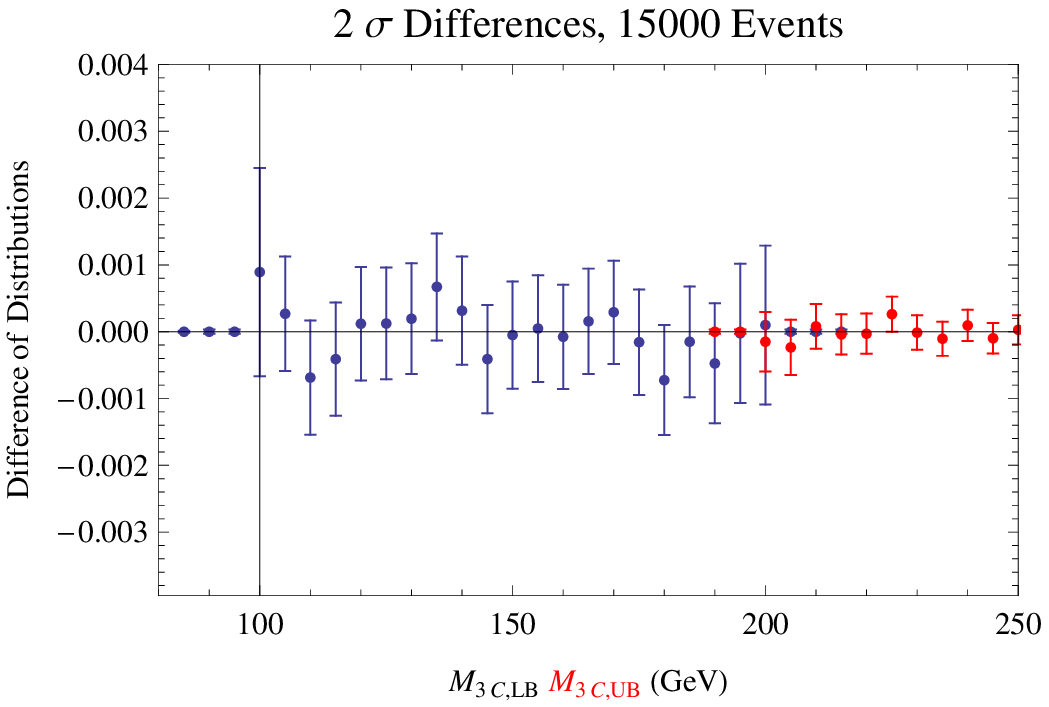}}
\caption{\label{FigM3CPdfDistributionDependence} The dependence of the $M_{3C}$ distributions on the parton collision energy.  The solid line shows the collision distributed according to Eq.~(\ref{EqSdep}), and the dashed line shows the collision energy fixed at $\sqrt{s}=600$ GeV. }
\end{figure}

Fig.~\ref{FigM3CPdfDistributionDependence} (Left) shows
the dependence of the $M_{3C}$ distributions on the parton collision energy.  The solid line shows the $M_{3C}$ distributions of events with collision energy $\sqrt{s}$ distributed according to
 \begin{equation}
 \rho(\sqrt{s}) = 12 \,M_{\N{2}}^2 \frac{\sqrt{s- 4\, M_{\N{2}}^2}}{s^2}, \label{EqSdep}
 \end{equation}
and the dashed line shows the $M_{3C}$ distributions of events with fixed $\sqrt{s}=600$ GeV.
Figure~\ref{FigM3CPdfDistributionDependence} (Right) shows the difference of these two distributions with $2 \sigma$ error bars as calculated from $15000$ events.  The two distributions are equal to within this numerical precision.

\begin{itemize}
\item {\bf{Effects of $\slashed{P}_T$ Cuts}}
\end{itemize}

\begin{figure}
\centerline{
\includegraphics[width=3.1in]{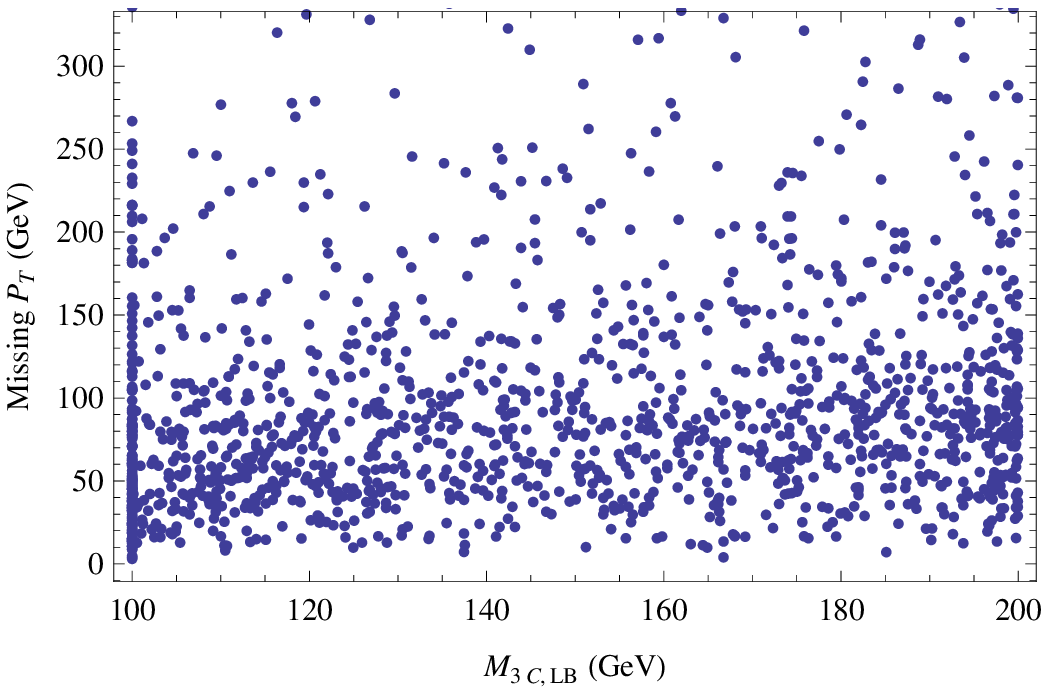}
\includegraphics[width=3.1in]{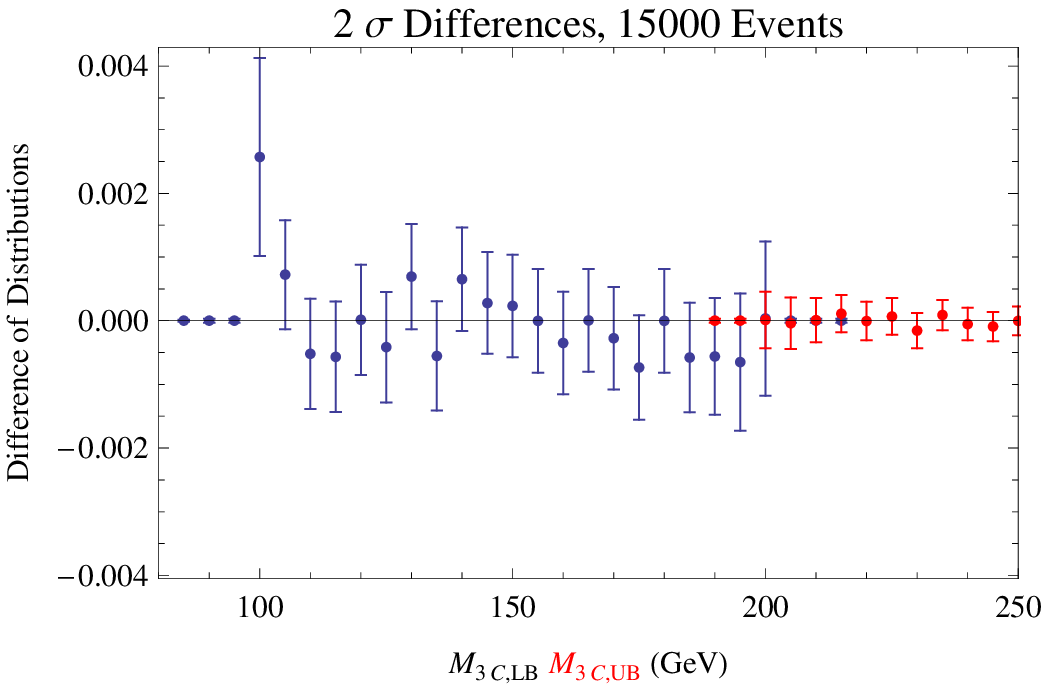}
}
\caption{\label{FigM3CPtCutDependenceDifference} The effect of missing transverse momentum
 cuts on the $M_{3C}$ distributions. (Left) The $M_{3C,LB}$ result versus the
 $\slashed{P}_T$.
(Right) The difference of the $M_{3C,UB}$ and $M_{3C,LB}$ distributions with and without the cut $\slashed{P}_T > 20$ GeV.
The smallest bins of $M_{3C,LB}$ are the only bins to be
statistically significantly affected. }
\end{figure}

As described in \cite{Ghosh:1999ix,Bisset:2005rn,Ross:2007rm,Barr:2008ba}, the standard-model four-lepton events with missing transverse momentum backgrounds are very strongly suppressed after a missing transverse momentum cut.
This requires an analysis of what effect a $\slashed{P}_T > 20$ GeV cut will have on the distribution shape.
Figure~\ref{FigM3CPtCutDependenceDifference} shows that the effect of this cut is significant dominantly at small $M_{3C}$.
On the left we see the $M_{3C,LB}$ result versus the $\slashed{P}_T$.
Unlike the $M_{2C}$ case in Ref~\cite{Barr:2008ba}, the $M_{3C}$ solutions in Fig.~\ref{FigM3CPtCutDependenceDifference} do not correlate with the $\slashed{P}_T$.
The right panel of Fig.~\ref{FigM3CPtCutDependenceDifference} shows the difference between the $M_{3C,UB}$ and $M_{3C,LB}$ distributions with and without the cut $\slashed{P}_T > 20$ GeV.
The smallest bins of $M_{3C,LB}$ are the only bins to be
statistically significantly affected.
The left-side suggests this lack of dependence on $\slashed{P}_T$ cuts is somewhat accidental and is due to the nearly uniform distribution of $M_{3C}$ solutions being removed by the cut.
The stronger dependence of the smallest $M_{3C}$ bins on the $\slashed{P}_T$ cut means we can either model the effect or exclude the first bins (about $10$ GeV worth) from the distribution used to predict the mass.  We will choose the latter because we will find that the background events also congregate in these first several bins.

\begin{itemize}
\item {\bf{Spin correlations}}
\end{itemize}

In our simulation to produce the ideal curves, we assumed each decay was uncorrelated with its spin in the rest frame of the decaying particle.
Spin correlations at production may affect this. However, such spin correlations are washed out when each branch of our assumed topology is at the end of longer decay chain. These upstream decays are the source of considerable UTM.

Some spin-correlation information can be easily taken into account. The $m_{12}$ (or $m_{34}$) distribution's shape is sensitive to the spin correlations along the decay chain \cite{Barr:2004ze}\cite{Athanasiou:2006ef}.  The observed $m_{12}$ (or $m_{34}$) distribution  can be used as an input to producing the ideal distribution shape.  In this way spin correlations along the decay chain can be taken into account in the simulations of the ideal distributions.

\begin{figure}
\centerline{\includegraphics[width=3.2in]{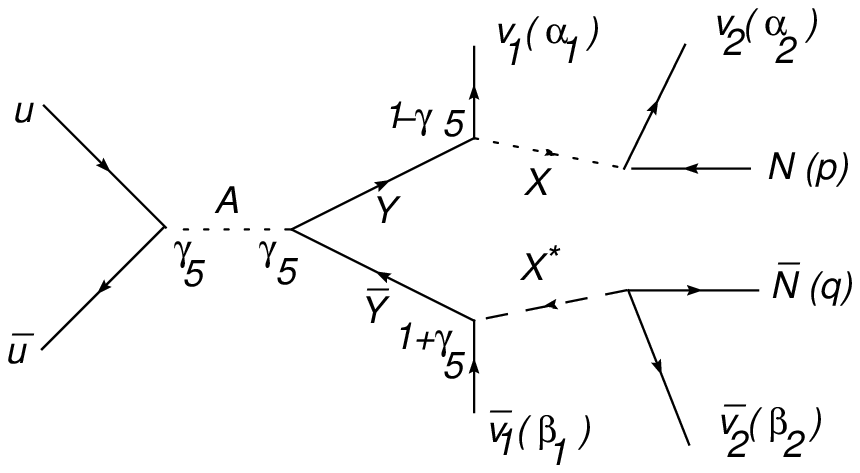}
\includegraphics[width=3.2in]{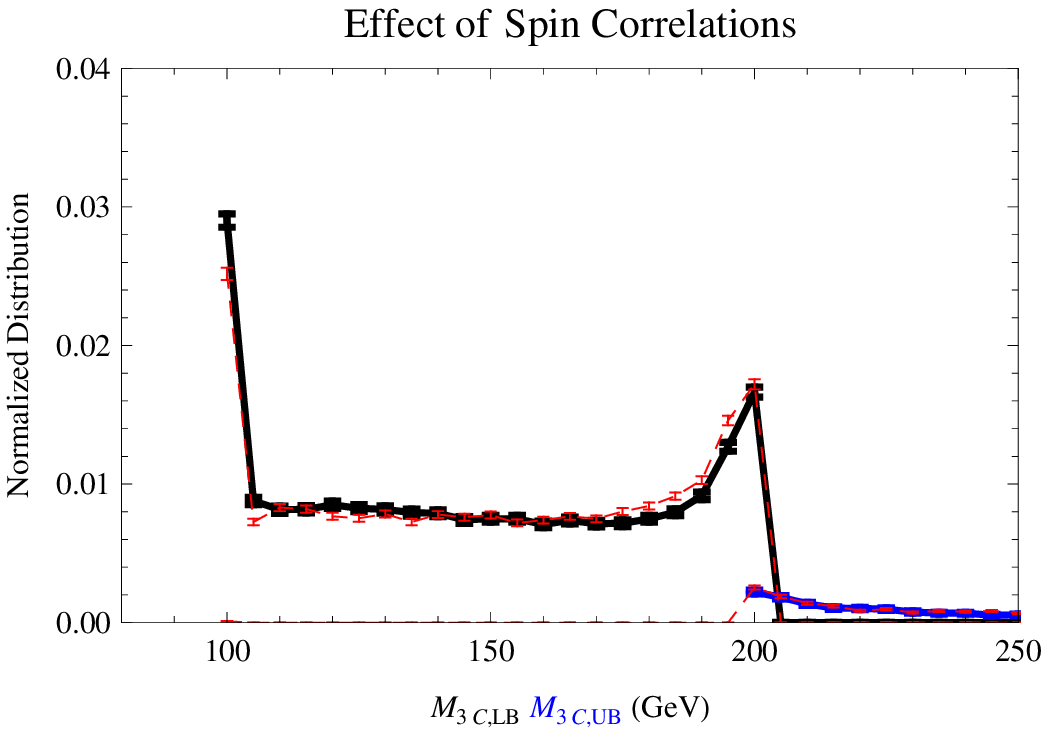}}
\caption{\label{FigM3CSpinCorrelations}Effect of a spin correlated process on the $M_{3C}$ distributions. Modeled masses are $M_Y=200$ GeV, $M_X=150$ GeV, and $M_N=100$ GeV.  The thick black and thick blue lines show the distributions of the uncorrelated lower-bound and upper-bound $M_{3C}$. The dotted red lines show the distributions of the spin correlated process.}
\end{figure}

Spin correlations between the two branches can also affect the distribution shape.
To demonstrate this we modeled a strongly spin-correlated direct production process.
Figure~\ref{FigM3CSpinCorrelations} (Left) shows the spin-correlated process that we consider.
Figure~\ref{FigM3CSpinCorrelations} (Right) shows the $M_{3C}$ upper-bound and lower-bound distributions from this process compared to the $M_{3C}$ distribution from the same topology and masses but without spin correlations.  We compare distributions with perfect energy resolution, $m_{v_1}=m_{v_2}=0$ GeV, $M_Y=200$ GeV, $M_X=150$ GeV, and $M_N=100$ GeV.  Our maximally spin-correlated process involves pair production of $Y$ through a pseudoscalar $A$.  The fermion $Y$ in both branches decays to a complex scalar $X$ and visible fermion $v_1$ through a purely chiral interaction. The scalar $X$ then decays to the dark-matter particle $N$ and another visible particle $v_2$.
The production of the pseudoscalar ensures that the $Y$ and $\bar{Y}$ are in a configuration $\sqrt{2}^{\,-1}(|\uparrow\downarrow\rangle + |\downarrow\uparrow\rangle)$.
The particle $Y$ then decays with $X$ preferentially aligned with the spin. The $\bar{Y}$ decays with $X^*$ preferentially aligned against the spin.
Because $X$ is a scalar, the particle $N$ decays uniformly in all directions from the rest frame of $X$.
The correlated directions of $X$ causes the two sources of missing transverse momentum to be preferentially parallel.  The resulting greater magnitude of missing transverse momentum increases the cases where $M_{3C}$ has a solution closer to the end point.  For this reason the spin correlated distribution (red dotted distribution) is above the uncorrelated distribution (black thick lower-bound distribution and blue thick upper-bound distribution).
The upper-bound distribution is statistically identical after 25000 events.
The lower-bound distribution clearly has been changed, but not by very much compared to the other factors on which the distribution shape depends that are described in this section.
This is due to the subsequent decay of the $X$ particle which lessens the likelihood that the two $N$s will be parallel.
For the remainder of the paper we assume no such spin correlations are present.

\begin{itemize}
\item {\bf{Backgrounds}}
\end{itemize}

The standard model (SM) backgrounds for four-leptons and missing transverse momentum are studied in  \cite{Ghosh:1999ix,Bisset:2005rn}.
In two previous publications \cite{Ross:2007rm}\cite{Barr:2008ba} we
 summarized the SM backgrounds and the dominance of supersymmetry (SUSY) backgrounds
for this channel.
As was mentioned earlier, the SM backgrounds for four leptons with missing transverse momentum are very strongly suppressed after a missing transverse momentum cut.

To improve the quality of the fit, a model for backgrounds can be created based on assumptions about the origin of the events and wedge-box analysis like those described in Bisset \etal \cite{Bisset:2008hm} and references therein.
We performed such a model in the Ref.~\cite{Barr:2008ba} and found the distribution shape isolated the correct mass of $\N{1}$ to within $1$ GeV with versus without the background model.
In the studies of $M_{3C}$ and SPS 1a in this paper, the insensitivity to the background is again observed.
If the background is present but not modeled, we observe a somewhat lower fit quality than in the absence of background, but the shift in $M_{\N{1}}$ is less than $1$ GeV which is within the predicted uncertainty.
In the SPS 1a example studied in the next section, SUSY background events
form about $12\%$ of the events.
As such, we do not try to model the background in this $M_{3C}$ study.

\section{Estimated Performance}
\label{SecM3CPerformance}

With an understanding of the factors affecting the shapes of the $M_{3C}$ distributions, we combine all the influences together and consider the mass-determination performance.
We follow the same modeling and simulation procedures used in Ref~\cite{Barr:2008ba}
except now we include an on-shell intermediate state and calculate $M_{3C}$.
We use \herwig\ \cite{Corcella:2002jc,Moretti:2002eu,Marchesini:1991ch} to generate events according to the SPS 1a benchmark point \cite{Allanach:2002nj}.
This is an {mSUGRA} point with $m_o=100$ GeV, $m_{1/2} = 250$ GeV, $A_o = -100$ GeV,  $\tan \beta = 10$, and sign$(\mu)=+$.
We initially assume the modeled mass differences of $\Delta M_{YN}=80.8$ GeV and $\Delta M_{XN}=47.0$ GeV
have been previously measured and take them as exact.
We later show how the $M_{3C}$ distribution-shape changes with the $m_{ll}$ end point also can be used to solve for the two mass differences.

The \herwig\ simulations select the charged leptons ($e^\pm$ and $\mu^\pm$)
produced in the decay of heavy objects (SUSY particles and $W$ and $Z$ bosons)
for further study
provided they satisfy basic selection criteria on transverse momentum
($p_T>10$~GeV) and pseudorapdity ($|\eta|<2.5$).
Leptons coming from hadron decays are usually contained within hadronic jets and
so can be experimentally rejected with high efficiency using energy or track isolation criteria.
This latter category of leptons was therefore not used in this study.
The acceptance criterion used for the hadronic final state was $|\eta|<5$.
The detector energy resolution functions used are described in Sec.~\ref{sec:detector}.

Like $M_{2C}$, the $M_{3C}$ distributions can be well-predicted
from observations.
When we determine the masses based on distribution shapes, the larger the area difference between two distributions representing different masses, the more accurately and precisely we will be able to tell the difference.
Unfortunately, the $M_{3C}$ distribution is sensitive to finite momentum-resolution errors and combinatoric errors which have the effect of decreasing the large area difference between the distributions of two different masses shown in Fig.~\ref{FigM3CIdeal150100}.

Just as in Ref~\cite{Barr:2008ba}, we model the distribution shape with a simple Mathematica Monte Carlo, and compare the predicted distribution shapes
against the \herwig\ data modeling the benchmark point SPS 1a.
We again use the observed UTM as an input to the Mathematica simulated ideal distributions.
By modeling with Mathematica, which does not use SUSY cross sections, and comparing to more realistic \herwig-generated data,
we hope to test that we understand the major
dependencies of the shape of the $M_{3C}$ distributions.
The Mathematica event generator produces events based on assumptions of a uniform angular distribution of the parents in the center-of-mass frame, the parent particles decay with a uniform angular distribution in the rest frame of the parent.  The particles are all taken to be on shell. The effect of $k_T>0$ is simulated by boosting the event in the transverse plane to compensate a specified $k_T$.

\begin{figure}
\centerline{
\includegraphics[width=3.1in]{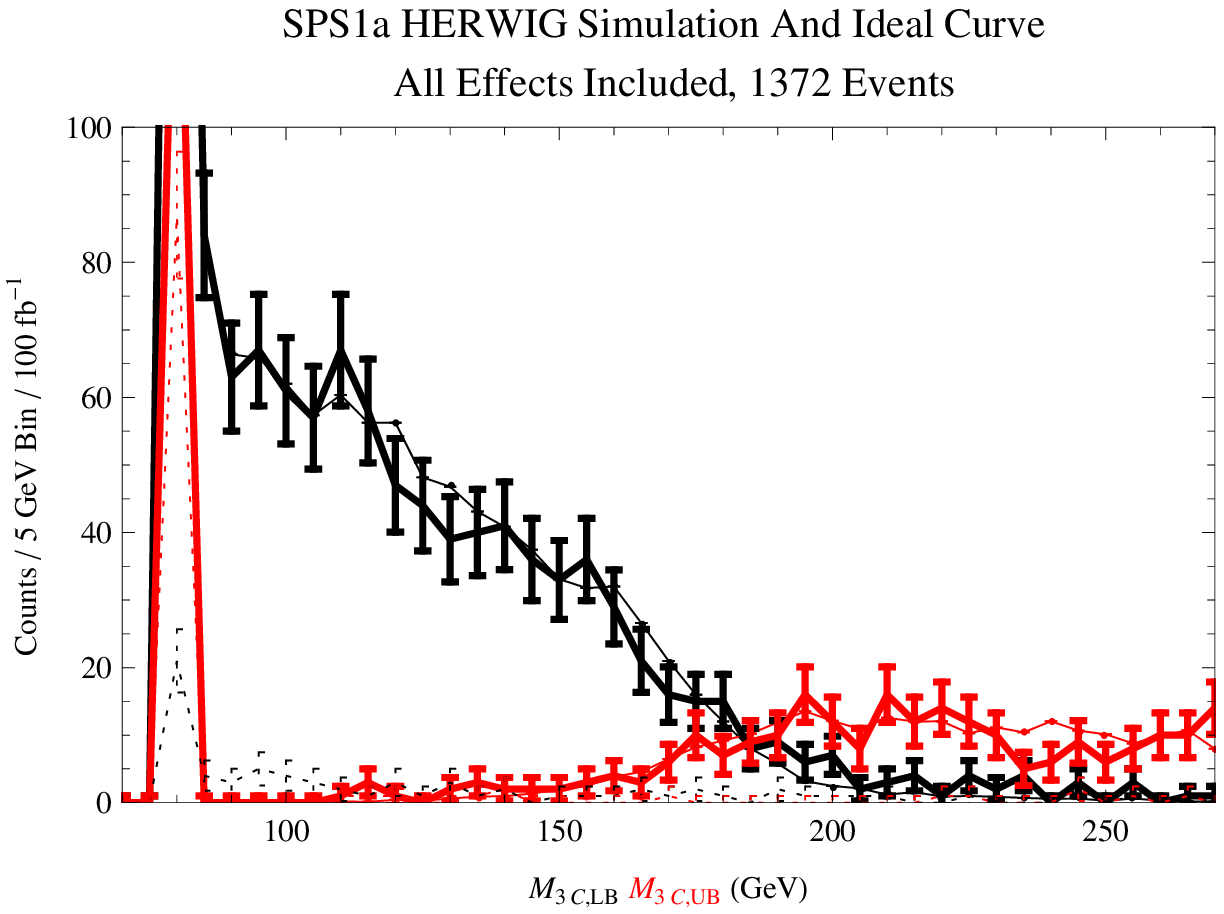}
\includegraphics[width=3.1in]{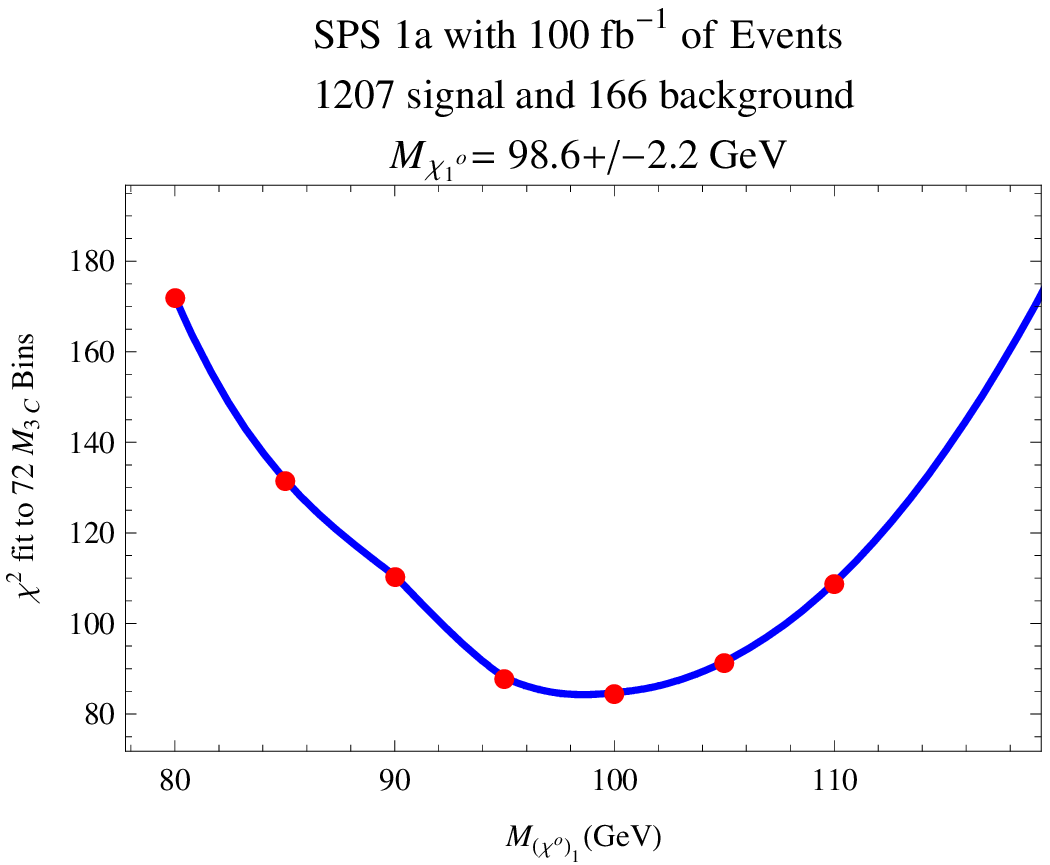}}
\caption{\label{FigM3CFitAllEffects} Fit of ideal $M_{3C}(M_{\N{1}})$ distributions to the \herwig\ generated $M_{3C}$ distributions. Includes combinatoric errors, backgrounds, energy resolution, and $\slashed{P}_T$ cuts. (Left) The observed \herwig\ counts versus the expected counts for ideal $M_{\N{1}}=95$ GeV.  (Right) The $\chi^2$ fit to ideal distributions of $M_{\N{1}}=80, 85, 90, 95, 100, 105, 110$ GeV. The correct mass in \herwig\ is $M_{\N{1}}= 96.0$ GeV.}
\end{figure}

The results are shown in Figure~\ref{FigM3CFitAllEffects}.
The left side of Fig.~\ref{FigM3CFitAllEffects} shows the $M_{3C}$ lower-bound
and upper-bound counts per $5$ GeV bin from the \herwig\ generated data, and it shows the predicted ideal counts calculated with Mathematica using the observed UTM distribution and assuming $M_{\N{1}}=95$ GeV.
The upper-bound and lower-bound distribution show very close agreement.
The background events are shown in dotted lines and are seen accumulating in the first few bins.  These are the same bins dominantly affected by $\slashed{P}_T$ cuts. For this reason we excluded these first two bins from the distribution fit.
The right side of Fig.~\ref{FigM3CFitAllEffects} shows the $\chi^2$ fit of the \herwig\ simulated data $M_{3C}$ distribution to the ideal $M_{3C}(M_{\N{1}})$ distribution with $M_{\N{1}}$ taken as the independent variable.
Ideal distribution shapes are calculated at values of $M_{\N{1}} = 80, 85, 90, 95, 100, 105, 110$ GeV.  The $\chi^2$ fitting procedure is described in more detail in appendix of Ref.~\cite{Barr:2008ba} except with $M_{2C}$ replaced with $M_{3C}$ and with the background fraction  $\lambda$ of the ideal distribution being tested against  fixed at $0$.
All effects discussed in this paper are included: combinatoric errors, SUSY backgrounds, energy resolution, and $\slashed{P}_T$ cuts.  Our ideal curves were based on the Mathematica simulations with $25000$ events per ideal curve.
Despite the presence of backgrounds, the $\chi^2$ per degree of freedom (number of bins) is not much above $1$ per bin.

 The particular fit shown in Fig.~\ref{FigM3CFitAllEffects} gives $M_{\N{1}}=98.6 \pm 2.2$ GeV where we measure uncertainty by using the positions at which $\chi^2$ is increased by one.
We repeat the fitting procedure on nine independent data sets each with $\approx 100 \fb^{-1}$ of \herwig\ data ($\approx 1400$ events for each set).
The mean and standard deviation of these nine fits give
$M_{\N{1}} = 96.8 \pm 3.7$ GeV. After $300 \fb^{-1}$ one should expect a $\sqrt{3}$ improvement in the uncertainty giving $\pm 2.2 \GeV$.  The correct mass in \herwig\ is $M_{\N{1}}= 96.05$ GeV.

\begin{figure}
\centerline{\includegraphics[width=3.3in]{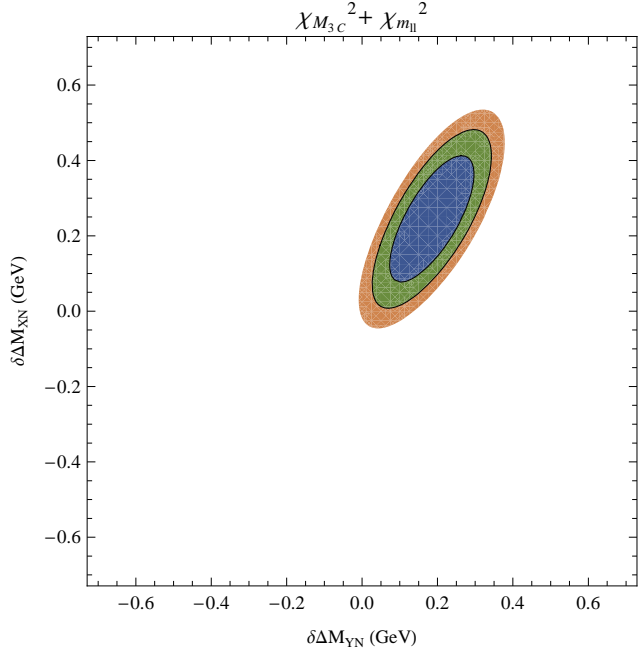}}
\caption{\label{FigM3CMllConstraint} Combined constraint from fitting both $\max m_{ll}$ and $M_{3C}$ with the mass difference as free parameters. We parameterized the difference from the true values in the model by $\Delta M_{YN}= 80.8 \GeV + \delta \Delta M_{YN}$ and $\Delta M_{XN}=47.0 \GeV + \delta \Delta M_{XN}$. We shown the $1, 2, 3 \sigma$ contours.}
\end{figure}

\section{Finding the mass differences using $m_{ll}$ edge and $M_{3C}$}
\label{SecMassDifferences}
Our technique also enables a combined fit to both the mass differences and the mass scale. The $m_{ll}$ end point in Eq.~(\ref{EqTwoBodyDecayEdge}) constrains a relationship between the three masses.
Gjelsten, Miller, and Osland estimate that this edge can be measured to better than $0.08 \GeV$  \cite{Gjelsten:2004ki,Gjelsten:2006tg}
using many different channels that
lead to the same edge, and after modeling energy resolution and background.
In the next several paragraphs we show that by combining this edge with the fits to the $M_{3C}$ upper-bound and lower-bound distribution shapes, we can constrain all three masses.

We first numerically calculate the effect of errors in the mass differences.
We use simulated data corresponding to $300 \fb^{-1}$ (about $3600$ signal and $450$ background events) including all the effects discussed.
We parameterize the error from the correct mass difference in the model by the variables $\delta \Delta M_{YN}$ and $\delta \Delta M_{XN}$ so that mass differences are given by $\Delta M_{YN}= 80.8 \GeV + \delta \Delta M_{YN}$ and $\Delta M_{XN}=47.0 \GeV + \delta \Delta M_{XN}$.
We calculate the $\chi^2_{M_{3C}}$ at $8$ points surrounding the correct mass difference by amounts $\delta \Delta M_{YN} = \pm 1 \GeV$ and $\delta \Delta M_{XN} = \pm 1 \GeV$.
The minimum $\chi^2_{M_{3C}}$ at each of the $9$ points gives the value of $M_{\N{1}}$ for each mass difference assumed.  The position of the minima can be parameterized by a quadratic near the true mass difference.  The resulting fit
 \begin{equation}
 M_{\N{1}}=96.4 + 1.9 \, (\delta \Delta M_{XN})^2 + 2.5  \,\delta \Delta M_{YN}\,
 \delta \Delta M_{XN}+3.2
   \delta \Delta M_{XN}-3.8 \,(\delta \Delta M_{YN})^2 - 8.3 \, \delta \Delta M_{YN}
   \label{EqMN1wDeltaMPropagated}
 \end{equation}
 shows in units of GeV how the mass $M_{\N{1}}$ is affected by small errors in the mass difference.

The $\chi^2_{M_{3C}}$ (for 72 bins and without a backgrounds model) at these $9$ different values for the mass difference provides another constraint on the mass differences.
Fitting the $\min \chi^2_{M_{3C}}$ to a general quadratic near the true mass difference gives
\begin{equation}
\min \chi^2_{M_{3C}} = 162 + 38 \, (\delta \Delta M_{XN})^2 - 8 \, \delta \Delta M_{YN} \delta \Delta M_{XN}
- 5 \,   \delta \Delta M_{XN}  - 25 \, \delta \Delta M_{YN} \label{EqChiSqM3C}.
\end{equation}
The $\min \chi^2_{M_{3C}}$ described by Eq.~(\ref{EqChiSqM3C}) shows a sloping valley.
The sides of the valley constrain $\delta \Delta M_{XN}$ as seen by
the large positive coefficient of $(\delta \Delta M_{XN})^2$.
The valley slopes downward along $\delta \Delta M_{YN}$ as can be seen by the large negative coefficient of $\delta \Delta M_{YN}$ which leaves this axis unbounded within the region studied.

The unconstrained direction along $\Delta M_{YN}$ can by constrained by the mass relationships given by  the end point $\max m_{ll}$ or by $M_{T2}$ as described in Ref.~\cite{Serna:2008zk}.  Here we work with $\max m_{ll}$ to provide this constraint.
We calculate the $\chi^2_{\max m_{ll}}$ using $\delta ( \max m_{ll}) = 0.08 \GeV$, and Eq.~(\ref{EqTwoBodyDecayEdge}) with $M_Y=\Delta M_{YN}+\delta \Delta M_{YN}+M_{\N{1}}$ and $M_X=\Delta M_{XN}+\delta \Delta M_{XN}+M_{\N{1}}$ where we use $M_{\N{1}}$ from Eq.~(\ref{EqMN1wDeltaMPropagated}).
This $\chi^2_{\max m_{ll}}$ constrains a diagonal path in $(\delta \Delta M_{YN},\delta \Delta M_{XN})$. The value of the $\chi^2_{\max m_{ll}}$ at the minimum is a constant along this path.
The combined constraint $\chi^2_{M_{3C}}$ to $\chi^2_{\max m_{ll}}$ leads to the
a minimum at $\delta \Delta M_{YN} = 0.18 \GeV$ and  $\delta \Delta M_{XN}= 0.25 \GeV$ where $M_{\N{1}}=95.7 \GeV$ as shown in Fig.~\ref{FigM3CMllConstraint}.
We have shown the contours where $\chi^2$ increases from its minimum by $1$,$2$ and $3$.
The uncertainty in the mass differences around this minimum is about $\pm 0.2 \GeV$.
The small bias from the true mass differences is due to the unconstrained $\chi^2_{M_{3C}}$ along $\delta \Delta M_{YN}$ and disappears with increasing statistics.
We can also use modeling to deduce the unbiased mass differences.

Putting together all effects and propagating the effects of uncertainty in the mass differences, we estimate a final performance of $M_{\N{1}}= 96.4 \pm 2.4$ GeV after $300 \fb^{-1}$ with about $3600$ signal events amid $450$ background events. We find the mass differences (without bias correction) of $ M_{\N{2}} - M_{\N{1}} = 81.0 \pm 0.2$ GeV and $M_{\tilde{l}_R} - M_{\N{1}}=44.3 \pm 0.2$ GeV.  This is to be compared to the  \herwig\ values of $M_{\N{1}}=96.0$ GeV, $M_{\N{2}} - M_{\N{1}} = 80.8$ GeV, and $M_{\tilde{l}_R} - M_{\N{1}}=44.3$ GeV.

\section{Discussion and Conclusions}
\label{SecM3CConclusions}

How does this performance compare to other techniques?
Because SPS 1a is commonly used as a test case, we can approximately compare performance with two different groups.
The technique of \cite{Bachacou:1999zb,Gjelsten:2004ki,Lester:2006yw,Gjelsten:2006tg}, which uses edges from cascade decays, determines the LSP mass to $\pm 3.4 \GeV$ with about five hundred thousand events from $300 \fb^{-1}$.  The approach of CEGHM \cite{Cheng:2008mg} assumes a pair of symmetric decay chains and assumes two events have the same topology and intermediate states.  They reach $\pm 2.8 \GeV$ using $700$ signal events after $300 \fb^{-1}$, but have a $2.5 \GeV$ systematic bias that needs modeling to remove.
Both techniques also constrain the mass differences.
By comparison we find $\pm 3.7$ GeV after $100 \fb^{-1}$ ($1200$ signal, $150$ background)
and estimate $\pm 2.4$ GeV after $300 \fb^{-1}$ ($3600$ signal, $450$ background) and propagating reasonable uncertainties in the mass differences.
The uncertainty calculations differ amongst research groups. Some groups estimate uncertainty from repeated trials, and others use the amount one can change the mass before $\chi^2$ increases by one.
Without careful comparison under like circumstances by the same research group, the optimal method is not clear.  What is clear is that fitting the $M_{3C,LB}$ and $M_{3C,UB}$ distributions determines the mass of
invisible particles as well if not better than the other known methods in both accuracy and precision.

In this paper, we have extended the constrained mass variable to the case with three new on-shell particle states.  We assume events with a symmetric, on-shell intermediate-state topology shown in Fig.~\ref{FigEventTopologyTovey}.
We can either assume that we have measured the mass difference between these new states through other techniques, or combine our technique with the $\max m_{ll}$ edge to find both mass differences and the mass scale.
The new constrained mass variables associated with events with these three new particle states are called
 $M_{3C,LB}$ and $M_{3C,UB}$, and they represent an event-by-event lower bound and upper bound (respectively) on the mass of the third lightest state possible while maintaining the constraints described in Eqs.~(\ref{EqM3Cc1})-(\ref{EqM3Cc7}).
We have shown that most of the $M_{2C}$ distribution properties described in the Refs~\cite{Ross:2007rm}\cite{Barr:2008ba} carry through to $M_{3C}$.
The additional particle state and mass difference enable a tighter event-by-event bound on the true mass.  The $M_{3C}$ distribution is more sensitive than the $M_{2C}$ distribution to the momentum and energy-resolution errors.  Studying the performance on the SPS 1a benchmark point, we find that despite the energy-resolution degradation, we are able to determine $M_{\N{1}}$ to \emph{at least} the same level of precision and accuracy  as that found by using cascade decays or by using other MSTs.

\section*{Acknowledgements}

We also want to thank Chris Lester and Giulia Zanderighi for helpful conversations and
comments on the manuscript and Laura Serna for reviewing the manuscript.
MS acknowledges support from
the United States Air Force Institute of Technology.
This work was partly supported by the Science and Technology Facilities Council of the United Kingdom.
The views expressed in this paper are those of the authors and do not reflect the official policy or
position of the United States Air Force, Department of Defense, or the US Government.

\appendix

\section{How to calculate $M_{3C}$}
\label{SecNumericallyCalculatingM3C}

To find the $M_{3C}$, we observe that if we assume masses of $Y$, $X$, and $N$ to be \footnote{We use $\chi$ to distinguish hypothetical masses $(\chi_Y,\chi_X,\chi_N)$ from the true masses $(M_Y,M_X,M_N)$.} $(\chi_Y,\chi_X,\chi_N)$  with the given mass differences then there are eight constraints
 \begin{eqnarray}
    (p+q)_T & = & \slashed{P}_T \\
    (\alpha_1+\alpha_2+p)^2 =(\beta_1+\beta_2+q)^2 & = & \chi^2_Y   \\
        (\alpha_2+p)^2=(\beta_2+q)^2  & = & \chi^2_{X} = (\chi_Y - \Delta M_{YN}+\Delta M_{XN})^2 \\
              p^2 = q^2 & = & \chi_N^2 = (\chi_Y - \Delta M_{YN})^2
 \end{eqnarray}
and eight unknowns, $p_\mu$ and $q_\mu$.
The spatial momenta $\vec{p}$ and $\vec{q}$ can be found as linear functions of
the $0^{\rm{th}}$ component of $p$ and $q$ by solving the matrix equation

\footnotesize
 \begin{eqnarray}
 \left(\begin{matrix} 1 & 0 & 0 & 1 & 0 & 0 \cr
                      0 & 1 & 0 & 0 & 1 & 0 \cr
                      -2 \alpha_x & -2 \alpha_y & -2\alpha_z & 0 & 0 & 0 \cr
   0 & 0 & 0 & -2 \beta_x & -2\beta_y & -2 \beta_z  \cr
                        -2 (\alpha_2)_x & -2 (\alpha_2)_y & -2(\alpha_2)_z & 0 & 0 & 0 \cr
   0 & 0 & 0 & -2 (\beta_2)_x & -2(\beta_2)_y & -2 (\beta_2)_z \end{matrix} \right)
 \left( \begin{matrix} p_x \cr p_y \cr p_z \cr q_x \cr q_y \cr q_z \end{matrix} \right)
 = \left( \begin{matrix}-(k+\alpha+\beta)_x \cr -(k+\alpha+\beta)_y \cr
   -2 \alpha_o p_o + (\chi_Y^2 - \chi_N^2) - \alpha^2 \cr
   -2 \beta_o q_o + (\chi_Y^2 - \chi_N^2) -\beta^2 \cr
   -2 (\alpha_2)_o p_o + (\chi_X^2-\chi_N^2) -(\alpha_2)^2 \cr
   -2 (\beta_2)_o q_o +  (\chi_X^2-\chi_N^2) - (\beta_2)^2 \end{matrix} \right)
\label{EqMatrixSolForpqVec}
 \end{eqnarray}
\normalsize

\noindent
where $\alpha=\alpha_1+\alpha_2$ and $\beta=\beta_1+\beta_2$.
We substitute $\vec{p}$ and $\vec{q}$ into the on-shell constraints
 \begin{eqnarray}
   p_o^2 - (\vec{p}(p_o,q_o))^2 = \chi_N^2 \label{Eqpo} \\
   q_o^2 - (\vec{q}(p_o,q_o))^2 = \chi_N^2 \label{Eqqo}
 \end{eqnarray}
giving two quadratic equations for $p_o$ and $q_o$.
These give four complex solutions for the pair $p_o$ and $q_o$.
We test each event for compatibility with a hypothetical triplet of masses $(\chi_Y,\chi_X,\chi_N)=(\chi_Y,\chi_Y-\Delta M_{YN}+\Delta M_{XN},\chi_Y-\Delta M_{YX})$.
If there are any purely real physical solutions where $p_o>0$ and $q_o>0$, then we consider the mass triplet $(\chi_Y,\chi_X,\chi_N)$ viable.

As we scan $\chi_Y$ while keeping the mass differences fixed, a solution begins to exist at a value less than or equal to $M_Y$ and then sometimes ceases to be a solution above $M_Y$. Sometimes there are multiple islands of solutions.
To find the $M_{3C}$, we can test each bin starting at $\chi_Y=\Delta M_{YN}$ along the path parameterized by $\chi_Y$ and the mass differences to find the first bin where at least one physical solution exists.
This is the lower-bound value of $M_{3C}$ for the event.

Likewise for an upper bound.  We begin testing at the largest conceivable mass scale we expect for the $Y$ particle state.  If a solution exists, we declare this a trivial $M_{3C,UB}$.  If no solution exists, then we search downward in mass scale until a solution exists.

A faster algorithm involves a bisection search for a solution within the window that starts at $\Delta M_{YN}$ and ends at our highest conceivable mass. We then use a binary search algorithm to find at what $\chi_Y$ the solution first appears for $M_{3C,LB}$ or at what $\chi_Y$ the solution disappears giving $M_{3C,UB}$. There are rare events where there are multiple islands of solutions. This occurs in about $0.01\%$ of the events with $0$ UTM and in about $0.1\%$ for $k_T = 250$ GeV. In our algorithm we neglect windows of solutions more narrow than $15$ GeV.  We report the smallest edge of the lower-mass island as the lower and the upper edge of the larger-mass island as the upper bound. Because of the presence of islands, we are not guaranteed that for an event solutions exist everywhere between $M_{3C,LB}$ and $M_{3C,UB}$.
With the inclusion of energy-resolution errors and background events, we also find cases where there are no solutions anywhere along the path being parameterized. If there is no solutions anywhere in the domain we make $M_{3C,LB}$ to be the largest conceivable mass scale, and we set $M_{3C,UB}=\Delta M_{YN}$.

\section{Uniqueness of Event Reconstruction}
\label{AppendixUniquenessOfReconstruction}


In Refs.~\cite{Ross:2007rm}\cite{Barr:2008ba} and in Sec.~\ref{SecM3CIntro} we claim that the events near an end point of $M_{2C}$ and $M_{3C}$ distributions (events that nearly saturate the bound) are nearly reconstructed.  This appendix offers a proof of the claim.  To prove uniqueness, we need to establish that as $M_{3C}$ or $M_{2C}$ of an event (lower bound or upper bound) approach the end point of the distributions, the solutions with different values of $q$ and $p$ approach a common solution.

\begin{figure}
\centerline{\includegraphics{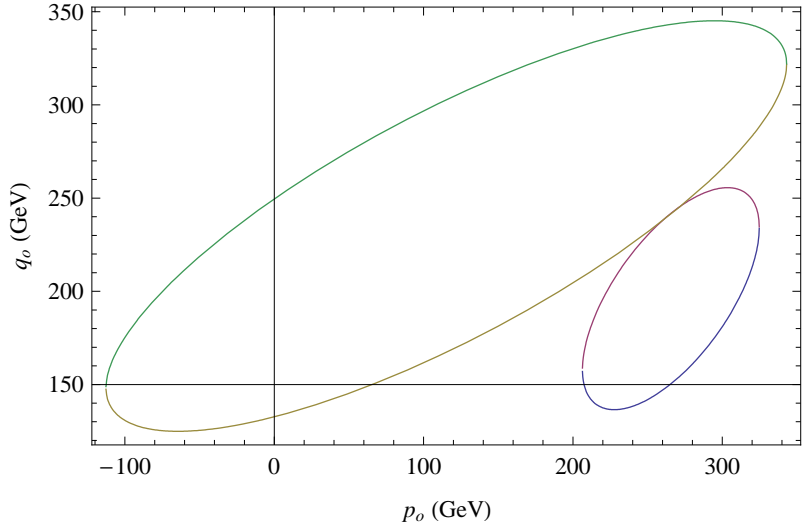}}
\caption{\label{FigpoqoEllipse} Shows the ellipses defined for $p_o$ and $q_o$ in  Eqs.~(\ref{Eqpo})-(\ref{Eqqo}) using the correct mass scale for an event that nearly saturates the $M_{3C}$ end point.  For this event, the $M_{3C}$ lies within $1\%$  of the end point and reconstructs $p$ and $q$ to within $4\%$. Perfect error resolution and combinatorics are assumed.}
\end{figure}
We begin with $M_{3C}$.  Section~\ref{SecNumericallyCalculatingM3C} shows that there are at most four solutions given $M_N$, $M_X$ and $M_Y$ formed by the intersection of two ellipses in $(p_o, q_o)$ defined by Eqs.~(\ref{Eqpo})-(\ref{Eqqo}) as shown in Fig.~\ref{FigpoqoEllipse}.   Consider the case that an event has a lower bound $M_{3C}$ near $M_Y$.  We are guaranteed that a solution occurs at the true mass scale when we choose the correct combinatoric assignments.  The ellipses either have a discrete number of solutions or they describe the same ellipse (an unlikely case which we dismiss as not relevant to realistic events).  As one varies the mass scale downward, the two ellipses drift and change shape and size so that four solutions become two solutions and eventually, at the value of $M_{3C}$ for the event, become one single solution.
When the disconnection of the two ellipses occurs near the true mass scale, the value of $M_{3C}$ will be near the end point.  The unique solutions for $p$ and $q$ given at $M_{3C}$  are nearly degenerate with the true values of $p$ and $q$ found when one uses the true masses to solve for $p$ and $q$.  The closer $M_{3C}$ is to the endpoint the closer the two ellipses are to intersecting at a single point when the true masses are used and to giving a unique reconstruction.  The example pictured in Fig.~\ref{FigpoqoEllipse} show an event with $M_{3C}$ within $1\%$ of the end point and where the $p$ and $q$ are reconstructed to within $4\%$.  This shows that for $M_{3C}$ events that are near the end point allowing for any choice of combinatorics then nearly reconstruct the true values for $p$ and $q$.   If there are combinatoric ambiguities, one need to test all combinatoric possibilities.  If the minimum combinatoric option has a lower bound at the end point, the above arguments follow unchanged.
The above arguments can be repeated to show $M_{3C,UB}$ near the end point also reconstructs the correct $p$ and $q$.

Next we turn to $M_{2C}$. For every event the lower-bounds satisfy $M_{2C}(M_Y-M_N) \leq M_{3C}(M_Y-M_N,M_X-M_N)$.  With $M_{2C}$ the propagator $(p+\alpha_2)^2$, which we can equate with $\chi_X^2$, is not fixed.  The kinematically allowed values for $\chi_X$ are $M_N^2 < \chi^2_X < M_Y^2$ assuming the visible states $\alpha_1$ and $\alpha_2$ are massless.
Eq.~(\ref{EqMatrixSolForpqVec}) shows that $\vec{p}$ and $\vec{q}$ solutions are linear in $\chi^2_X$ with no terms dependent on $\chi_X$ alone or other powers of $\chi_X$.  Including $\chi_X^2$ as a free parameter in Eqs.~(\ref{Eqpo})-(\ref{Eqqo}) leads to two ellipsoids (or hyperboloids) in the space $(p_o,q_o,\chi_X^2)$.  We will assume without loss of generality that these are ellipsoids because the arguments follow unchanged if they are hyperboloids.  Again, at the true mass scale we are guaranteed the two ellipsoids intersect at an ellipse.  We again neglect the physically unlikely case the two ellipsoids are degenerate.  Now as one varies the mass scale the two ellipsoids drift and change shape and size.  The $M_{2C}$ value then corresponds to the mass scale where the two ellipsoids are in contact at one point.
As we select events with a value of $M_{2C}$ that approaches the true mass scale the intersection of the two ellipsoids shrink to a point giving a unique reconstruction of $p$ and $q$.
The combinatoric ambiguities for $M_{2C}$ are avoided by selecting events with two distinct OSSF pairs.
Events that saturate the upper bound of $M_{2C}$ also reconstruct $p$ and $q$ by the same logic as above.


\begin{thebibliography}{10}

\bibitem{Ross:2007rm}
G.~G. Ross and M.~Serna, {\it {Mass Determination of New States at Hadron
  Colliders}},  {\em Phys. Lett.} {\bf B665} (2008) 212--218,
  [\href{http://xxx.lanl.gov/abs/0712.0943}{{\tt 0712.0943}}].

\bibitem{Barr:2008ba}
A.~J. Barr, G.~G. Ross, and M.~Serna, {\it {The Precision Determination of
  Invisible-Particle Masses at the LHC}},  {\em Phys. Rev.} {\bf D78} (2008)
  056006, [\href{http://xxx.lanl.gov/abs/0806.3224}{{\tt 0806.3224}}].

\bibitem{Bachacou:1999zb}
H.~Bachacou, I.~Hinchliffe, and F.~E. Paige, {\it Measurements of masses in
  sugra models at lhc},  {\em Phys. Rev.} {\bf D62} (2000) 015009,
  [\href{http://xxx.lanl.gov/abs/hep-ph/9907518}{{\tt hep-ph/9907518}}].

\bibitem{phdthesis-lester}
C.~Lester, {\em Model independent sparticle mass measurements at {ATLAS}}.
\newblock {PhD} dissertation, University of Cambridge, Department of Physics,
  December, 2001.
\newblock {CERN-THESIS-2004-003}.

\bibitem{Allanach:2000kt}
B.~C. Allanach, C.~G. Lester, M.~A. Parker, and B.~R. Webber, {\it {Measuring
  sparticle masses in non-universal string inspired models at the LHC}},  {\em
  JHEP} {\bf 09} (2000) 004,
  [\href{http://xxx.lanl.gov/abs/hep-ph/0007009}{{\tt hep-ph/0007009}}].

\bibitem{Gjelsten:2004ki}
B.~K. Gjelsten, D.~J. Miller, and P.~Osland, {\it {Measurement of SUSY masses
  via cascade decays for SPS 1a}},  {\em JHEP} {\bf 12} (2004) 003,
  [\href{http://xxx.lanl.gov/abs/hep-ph/0410303}{{\tt hep-ph/0410303}}].

\bibitem{Bisset:2008hm}
M.~Bisset, N.~Kersting, and R.~Lu, {\it {Improving SUSY Spectrum Determinations
  at the LHC with Wedgebox and Hidden Threshold Techniques}},
  \href{http://xxx.lanl.gov/abs/0806.2492}{{\tt 0806.2492}}.

\bibitem{Lester:2006yw}
C.~G. Lester, {\it Constrained invariant mass distributions in cascade decays:
  The shape of the 'm(qll)-threshold' and similar distributions},  {\em Phys.
  Lett.} {\bf B655} (2007) 39--44,
  [\href{http://xxx.lanl.gov/abs/hep-ph/0603171}{{\tt hep-ph/0603171}}].

\bibitem{Gjelsten:2006tg}
B.~K. Gjelsten, D.~J. Miller, P.~Osland, and A.~R. Raklev, {\it Mass
  determination in cascade decays using shape formulas},  {\em AIP Conf. Proc.}
  {\bf 903} (2007) 257--260,
  [\href{http://xxx.lanl.gov/abs/hep-ph/0611259}{{\tt hep-ph/0611259}}].

\bibitem{Cheng:2007xv}
H.-C. Cheng, J.~F. Gunion, Z.~Han, G.~Marandella, and B.~McElrath, {\it {Mass
  Determination in SUSY-like Events with Missing Energy}},  {\em JHEP} {\bf 12}
  (2007) 076, [\href{http://xxx.lanl.gov/abs/0707.0030}{{\tt 0707.0030}}].

\bibitem{Nojiri:2007pq}
M.~M. Nojiri, G.~Polesello, and D.~R. Tovey, {\it {A hybrid method for
  determining SUSY particle masses at the LHC with fully identified cascade
  decays}},  {\em JHEP} {\bf 05} (2008) 014,
  [\href{http://xxx.lanl.gov/abs/0712.2718}{{\tt 0712.2718}}].

\bibitem{Cheng:2008mg}
H.-C. Cheng, D.~Engelhardt, J.~F. Gunion, Z.~Han, and B.~McElrath, {\it
  {Accurate Mass Determinations in Decay Chains with Missing Energy}},  {\em
  Phys. Rev. Lett.} {\bf 100} (2008) 252001,
  [\href{http://xxx.lanl.gov/abs/0802.4290}{{\tt 0802.4290}}].

\bibitem{Cho:2007qv}
W.~S. Cho, K.~Choi, Y.~G. Kim, and C.~B. Park, {\it {Gluino Stransverse Mass}},
   {\em Phys. Rev. Lett.} {\bf 100} (2008) 171801,
  [\href{http://xxx.lanl.gov/abs/0709.0288}{{\tt 0709.0288}}].

\bibitem{Cho:2007dh}
W.~S. Cho, K.~Choi, Y.~G. Kim, and C.~B. Park, {\it {Measuring superparticle
  masses at hadron collider using the transverse mass kink}},  {\em JHEP} {\bf
  02} (2008) 035, [\href{http://xxx.lanl.gov/abs/0711.4526}{{\tt 0711.4526}}].

\bibitem{Barr:2007hy}
A.~J. Barr, B.~Gripaios, and C.~G. Lester, {\it {Weighing Wimps with Kinks at
  Colliders: Invisible Particle Mass Measurements from Endpoints}},  {\em JHEP}
  {\bf 02} (2008) 014, [\href{http://xxx.lanl.gov/abs/0711.4008}{{\tt
  0711.4008}}].

\bibitem{Allanach:2002nj}
B.~C. Allanach {\em et.~al.}, {\it The snowmass points and slopes: Benchmarks
  for susy searches},  \href{http://xxx.lanl.gov/abs/hep-ph/0202233}{{\tt
  hep-ph/0202233}}.

\bibitem{Serna:2008zk}
M.~Serna, {\it {A short comparison between $m_{T2}$ and $m_{CT}$}},  {\em JHEP}
  {\bf 06} (2008) 004, [\href{http://xxx.lanl.gov/abs/0804.3344}{{\tt
  0804.3344}}].

\bibitem{Cho:2008tj}
W.~S. Cho, K.~Choi, Y.~G. Kim, and C.~B. Park, {\it {$M_{T2}$-assisted on-shell
  reconstruction of missing momenta and its application to spin measurement at
  the LHC}},  \href{http://xxx.lanl.gov/abs/0810.4853}{{\tt 0810.4853}}.

\bibitem{Lester:1999tx}
C.~G. Lester and D.~J. Summers, {\it Measuring masses of semi-invisibly
  decaying particles pair produced at hadron colliders},  {\em Phys. Lett.}
  {\bf B463} (1999) 99--103,
  [\href{http://xxx.lanl.gov/abs/hep-ph/9906349}{{\tt hep-ph/9906349}}].

\bibitem{Barr:2003rg}
A.~Barr, C.~Lester, and P.~Stephens, {\it {m(T2): The truth behind the
  glamour}},  {\em J. Phys.} {\bf G29} (2003) 2343--2363,
  [\href{http://xxx.lanl.gov/abs/hep-ph/0304226}{{\tt hep-ph/0304226}}].

\bibitem{Cheng:2008hk}
H.-C. Cheng and Z.~Han, {\it {Minimal Kinematic Constraints and $M_{T2}$}},
  \href{http://xxx.lanl.gov/abs/0810.5178}{{\tt 0810.5178}}.

\bibitem{Burns:2008va}
M.~Burns, K.~Kong, K.~T. Matchev, and M.~Park, {\it {Using Subsystem $M_{T2}$
  for Complete Mass Determinations in Decay Chains with Missing Energy at
  Hadron Colliders}},  \href{http://xxx.lanl.gov/abs/0810.5576}{{\tt
  0810.5576}}.

\bibitem{Goldstein:1993mj}
G.~R. Goldstein, K.~Sliwa, and R.~H. Dalitz, {\it {Observing top-quark
  production at the Fermilab Tevatron}},  {\em Phys. Rev.} {\bf 47} (1993)
  967--972.

\bibitem{Kondo:1993in}
K.~Kondo, T.~Chikamatsu, and S.~H. Kim, {\it {Dynamical likelihood method for
  reconstruction of events with missing momentum. 3: Analysis of a CDF high
  p(T) e mu event as t anti-t production}},  {\em J. Phys. Soc. Jap.} {\bf 62}
  (1993) 1177--1182.

\bibitem{Raja:1996vz}
R.~Raja, {\it {On measuring the top quark mass using the dilepton decay
  modes}},  {\em ECONF} {\bf C960625} (1996) STC122,
  [\href{http://xxx.lanl.gov/abs/hep-ex/9609016}{{\tt hep-ex/9609016}}].

\bibitem{Raja:1997qs}
R.~Raja, {\it {Remark on the errors associated with the Dalitz-Goldstein
  method}},  {\em Phys. Rev.} {\bf D56} (1997) 7465--7465.

\bibitem{Brandt:2006uc}
O.~Brandt, {\it {Measurement of the mass of the top quark in dilepton final
  states with the D0 detector}}, . FERMILAB-MASTERS-2006-03.

\bibitem{AtlasTDR}
A.~Collaboration, {\it {ATLAS} computing : Technical design report},  {\em
  {CERN}, {ATLAS-TDR-017},{CERN-LHCC-2005-022}} (2005).

\bibitem{CMSTDR}
C.~Collaboration, {\em CMS physics : Technical Design Report}.
\newblock CERN Report No: CERN-LHCC-2006-001 ; CMS-TDR-008-1, 2006.

\bibitem{Ghosh:1999ix}
D.~K. Ghosh, R.~M. Godbole, and S.~Raychaudhuri, {\it Signals for
  r-parity-violating supersymmetry at a 500-gev e+ e- collider},
  \href{http://xxx.lanl.gov/abs/hep-ph/9904233}{{\tt hep-ph/9904233}}.

\bibitem{Bisset:2005rn}
M.~Bisset, N.~Kersting, J.~Li, F.~Moortgat, and Q.~Xie, {\it Pair-produced
  heavy particle topologies: Mssm neutralino properties at the lhc from gluino
  / squark cascade decays},  {\em Eur. Phys. J.} {\bf C45} (2006) 477--492,
  [\href{http://xxx.lanl.gov/abs/hep-ph/0501157}{{\tt hep-ph/0501157}}].

\bibitem{Barr:2004ze}
A.~J. Barr, {\it {Using lepton charge asymmetry to investigate the spin of
  supersymmetric particles at the LHC}},  {\em Phys. Lett.} {\bf B596} (2004)
  205--212, [\href{http://xxx.lanl.gov/abs/hep-ph/0405052}{{\tt
  hep-ph/0405052}}].

\bibitem{Athanasiou:2006ef}
C.~Athanasiou, C.~G. Lester, J.~M. Smillie, and B.~R. Webber, {\it
  {Distinguishing spins in decay chains at the Large Hadron Collider}},  {\em
  JHEP} {\bf 08} (2006) 055,
  [\href{http://xxx.lanl.gov/abs/hep-ph/0605286}{{\tt hep-ph/0605286}}].

\bibitem{Corcella:2002jc}
G.~Corcella {\em et.~al.}, {\it {HERWIG 6.5 release note}},
  \href{http://xxx.lanl.gov/abs/hep-ph/0210213}{{\tt hep-ph/0210213}}.

\bibitem{Moretti:2002eu}
S.~Moretti, K.~Odagiri, P.~Richardson, M.~H. Seymour, and B.~R. Webber, {\it
  {Implementation of supersymmetric processes in the HERWIG event generator}},
  {\em JHEP} {\bf 04} (2002) 028,
  [\href{http://xxx.lanl.gov/abs/hep-ph/0204123}{{\tt hep-ph/0204123}}].

\bibitem{Marchesini:1991ch}
G.~Marchesini {\em et.~al.}, {\it {HERWIG: A Monte Carlo event generator for
  simulating hadron emission reactions with interfering gluons. Version 5.1 -
  April 1991}},  {\em Comput. Phys. Commun.} {\bf 67} (1992) 465--508.

\end{thebibliography}

\newcommand{\noopsort}[1]{} \newcommand{\printfirst}[2]{#1}
  \newcommand{\singleletter}[1]{#1} \newcommand{\switchargs}[2]{#2#1}
\providecommand{\href}[2]{#2}\begingroup\raggedright\endgroup

\end{document}